\documentclass[superscriptaddress, twocolumn, prx, aps, longbibliography, nofootinbib]{revtex4-1}

\usepackage{amsmath, amssymb, color, wasysym, esint}
\makeatletter
\newsavebox{\@brx}
\newcommand{\llangle}[1][]{\savebox{\@brx}{\(\m@th{#1\langle}\)}\mathopen{\copy\@brx\kern-0.5\wd\@brx\usebox{\@brx}}}
\newcommand{\rrangle}[1][]{\savebox{\@brx}{\(\m@th{#1\rangle}\)}%
  \mathclose{\copy\@brx\kern-0.5\wd\@brx\usebox{\@brx}}}
\makeatother

\makeatletter
\newsavebox{\@brxx}
\newcommand{\lllangle}[1][]{\savebox{\@brxx}{\(\m@th{#1\langle}\)}%
  \mathopen{\copy\@brxx\kern-0.5\wd\@brxx\usebox{\@brxx}\kern-0.5\wd\@brxx\usebox{\@brxx}}}
\newcommand{\rrrangle}[1][]{\savebox{\@brxx}{\(\m@th{#1\rangle}\)}%
  \mathclose{\copy\@brxx\kern-0.5\wd\@brxx\usebox{\@brxx}\kern-0.5\wd\@brxx\usebox{\@brxx}}}
\makeatother

\usepackage{graphicx}
\usepackage{bm}
\usepackage{xcolor}
\usepackage{xspace}
\usepackage{float}
\usepackage{lineno}

\definecolor{linkcolor}{rgb}{0,0,0.6} 
\usepackage[pdftex,colorlinks=true,
	pdfstartview = FitV,
	linkcolor    = linkcolor,
	citecolor    = linkcolor,
	urlcolor     = linkcolor,
	hyperindex   = true,
	hyperfigures = false]{hyperref}

\begin{document}

\title{Hydrodynamics of pulsating active liquids}

\author{Tirthankar Banerjee}
\email{tirthankar@niser.ac.in}
\affiliation{Department of Physics and Materials Science, University of Luxembourg, L-1511 Luxembourg City, Luxembourg}
\affiliation{School of Physical Sciences, National Institute of Science Education and Research, Jatani, Odisha, 752050 India}
\affiliation{Homi Bhaba National Institute, Training School Complex, Anushakti Nagar, Mumbai, 400094 India}

\author{Thibault Desaleux}
\affiliation{Department of Physics and Materials Science, University of Luxembourg, L-1511 Luxembourg City, Luxembourg}

\author{Jonas Ranft}
\email{jonas.ranft@ens.psl.eu}
\affiliation{Institut de Biologie de l’ENS, \'Ecole Normale Sup\'erieure, CNRS, Inserm, Universit\'e PSL, 46 rue d’Ulm, 75005 Paris, France}

\author{\'Etienne Fodor}
\email{etienne.fodor@uni.lu}
\affiliation{Department of Physics and Materials Science, University of Luxembourg, L-1511 Luxembourg City, Luxembourg}

\begin{abstract}
Inspired by dense contractile tissues, where cells are subject to periodic deformation, we formulate and study a generic hydrodynamic theory of pulsating active liquids. Combining mechanical and phenomenological arguments, we postulate that the mechanochemical feedback between the local phase, which describes how cells deform due to autonomous driving, and the local density can be described in terms of a free energy. Our hydrodynamics captures the three main states emerging in its particle-based counterparts: a globally cycling state, a homogeneous arrested state with constant phase, and a state with propagating radial waves. Remarkably, we show that the competition between these states can be rationalized intuitively in terms of an effective landscape, and argue that waves can be regarded as secondary instabilities. Linear stability analysis of the arrested and cycling states, including the role of fluctuations, leads to predictions for the phase boundaries. Overall, our results demonstrate that our minimal, yet non-trivial model provides a relevant platform to study the rich phenomenology of pulsating liquids.
\end{abstract}

\maketitle


\section{Introduction}\label{sec:intro}

Active matter encompasses a large class of nonequilibrium systems where individual components can autonomously consume energy to power some sustained dynamics~\cite{active-matter-review-bechinger, active-matter-marchetti, Marchetti2018}. The study of various active phenomena, such as flocking~\cite{vicsek-prl, toner-tu, chate-flocking-first-order, Chate2020}, motility-induced phase separation~\cite{MIPS}, and active turbulence~\cite{active-turbulence-yeomans, active-turbulence-alert}, has largely driven the development of nonequilibrium statistical physics in the last decades. While many studies have focused on dilute systems~\cite{Chate2020, MIPS}, others have been concerned with dense assemblies of active units, such as the acto-myosin cortex or biological tissues described as active gels~\cite{Kruse...Sekimoto2004, Joanny2009, Jonas2010, Prost2015}. The corresponding phenomenology features oscillations~\cite{martin2009pulsed, Hakim-oscillations-NatComm14, Peyret-oscillations-waves19} and wave propagation~\cite{Serra-Picamal-Natphys2012, Zaritsky-PLoS, Banerjee-Marchetti-PRL2015, tlili-RSoc18, Petrolli-PRL2019, Hino-DevCell2020, Armon-PNAS, Young1997}, associated with various biological functions~\cite{kruse2011spontaneous, collinet2021programmed}. Interestingly, there is a long history of continuum models to describe propagating electromechanical patterns in cardiac tissues~\cite{ALIEV1996293, Shajahan2009}, where transition between different kinds of patterns such as rotating spiral waves or defect turbulence, could be linked with different physiological disorders, namely tachycardia or ventricular fibrillation~\cite{karma-annrev, Nitsan-cardiomyocyte, christoph2018electromechanical}. As other examples, phase field models incorporating cell polarity, self-propulsion and mechanical coupling between cells have been employed to explain experimentally observed collective oscillations~\cite{Peyret-oscillations-waves19}, or models of oscillatory dynamics in electrically coupled uterine cells~\cite{Xu2015} that are believed to underlie uterine contractions during labor~\cite{Myers2017, Maeda2013UterineCI}.

Recently, minimal models of deformable units have led to a novel paradigm in soft and active matter~\cite{Peyret-oscillations-waves19,PhysRevX.6.021011,Manning-essay,PhysRevLett.129.148101,PhysRevLett.130.038202, goth2025, Zhang_2025}. A prominent example is the so-called vertex model, originally developed to  rationalize geometric properties of epithelial tissues in terms of their mechanics~\cite{farhadifar2007influence, staple2010mechanics}. More recent works considered versions of the model where internal parameters are dynamical variables, to account e.g.~for mechanochemical couplings between cells or tissue plasticity~\cite{Armon-Comm-phys, Soto2022, Hannezo-PRXLife, Shiladitya-arxiv23, Sadjad-Manning-arXiv}. Several of these studies showed how mechanical feedback at the single-cell level can lead to patterns of contraction pulses and waves in confluent tissues~\cite{Armon-Comm-phys, Soto2022, Hannezo-PRXLife, Shiladitya-arxiv23}. In an approach complementary to vertex models, recent studies on pulsating repulsive particles~\cite{Togashi2019, Yiwei-Etienne-PAM, Liu_2024, li2024fluidization, pineros2024biased, casagrande2026a, casagrande2026b}, whose sizes are subject to periodic changes, have shown how the interplay between deformation and synchronization yields dynamical patterns reminiscent of the contraction waves in pulsatile tissues~\cite{karma-annrev, collinet2021programmed}.

An important challenge is to propose a hydrodynamic description which accurately captures the collective properties of deforming particles. Several studies have proposed hydrodynamic equations for the vertex model when considering fixed reference area and perimeter for all cells~\cite{ishihara-pre2017, C8SM00446C, hernandez2021, grossman2022instabilities}. For pulsating particles, their hydrodynamics has been derived with a bottom-up approach by coarse-graining the microscopic model under the assumption that the local phase, which describes the size oscillation, does not couple to the local density~\cite{Yiwei-Etienne-PAM}. Therefore, although the corresponding hydrodynamics features interesting connections with the complex Ginzburg-Landau equation (CGLE)~\cite{aranson-review}, which is widely used in synchronization theory, it does not capture the emergence of density waves observed experimentally.

In short, a consistent hydrodynamics of pulsating particles which properly integrates the mechanochemical feedback between density and phase remains to be built. First, such a theory would allow one to compare more closely existing descriptions of confluent tissues, formulated in terms of continuum mechanics~\cite{Jonas2010, tlili2015colloquium, ishihara-pre2017,C8SM00446C, hernandez2021, grossman2022instabilities}, with models of pulsating particles~\cite{Togashi2019,Yiwei-Etienne-PAM,Liu_2024,li2024fluidization, pineros2024biased,manacorda2023pulsating}. Second, it would ideally bridge the gap between these continuum models, based on conservation laws and constitutive relations, and field theories of synchronisation that rely essentially on phenomenological laws~\cite{aranson-review, Ritort2005}. Therefore, our goal is to delineate the essential ingredients, combining mechanical and phenomenological arguments, to build a generic hydrodynamic theory that captures the phenomenology of a broad class of pulsating systems.

\begin{figure}      
    \includegraphics[width=0.99\columnwidth]{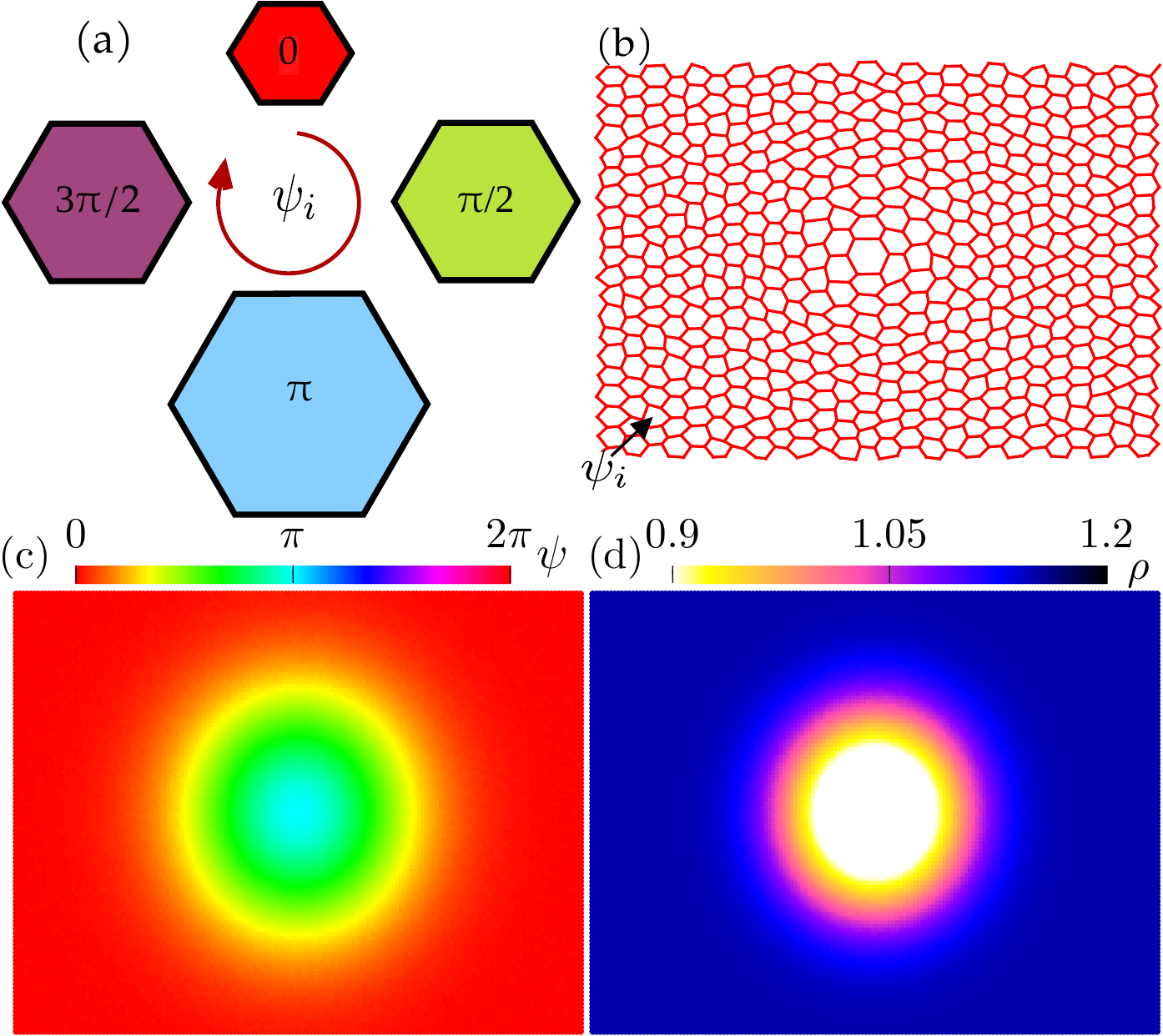}
    \caption{Schematic representation of pulsating liquids. For illustration, we consider here the case of a confluent tissue made of cells with a polygonal shape.
    (a)~Sketch of the cyclic change of a cell's preferred area described by the internal phase $\psi_i$. 
    (b)~A local increase in cell sizes, here at the center, always comes with a local decrease in density due to the confluent packing of cells. The corresponding coarse-grained fields of (c)~the phase $\psi({\bf r})$ and (d)~the density $\rho({\bf r})$ are assumed to vary smoothly over length scales large compared to the size of individual cells. As in the particle-based picture, a local increase in size ($\psi({\bf r})\simeq\pi$) is associated with a local decrease of density.
    }
\label{model-fig}
\end{figure}

\begin{figure}[b]
    \includegraphics[width=\columnwidth]{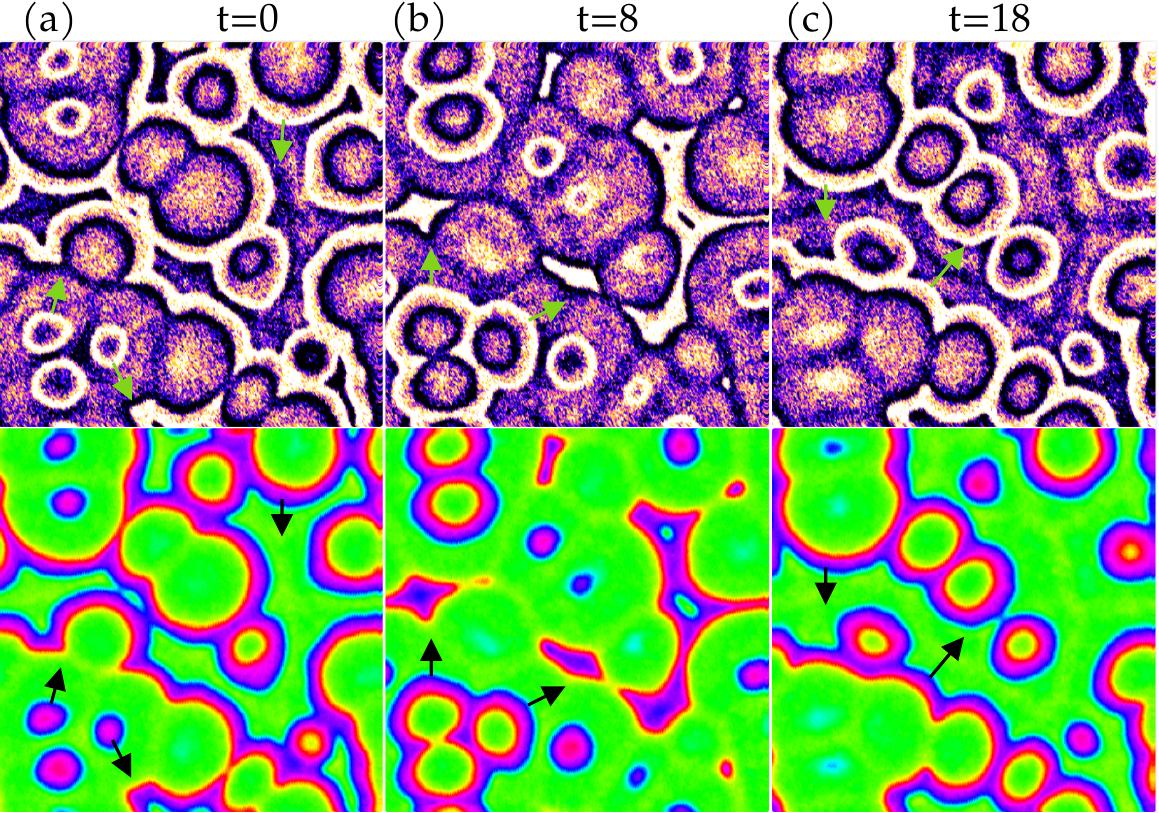}
    \caption{Pulsating radial waves propagate synchronously for density $\rho$ [top panel] and phase $\psi$ [bottom panel]~\cite{Movie1}. Variations of the local phase generate local stresses, which in turn produce large-scale flows and advect the density. Arrows indicate directions of propagation.
    Parameters: $\beta=9$, $\bar\rho=1.04$, $\alpha=1$, $D_\rho=0.01=D_\psi$, $\epsilon=0.5, ~dx=0.64,~ dt=0.001$. Colormaps as in Fig.~\ref{model-fig}.
    }
\label{model-fig1}
\end{figure}

In this article, we set out to formulate such a theory by explicitly describing the hydrodynamic coupling between density and phase. We assume that individual self-propulsion is negligible, so that activity enters purely through the drive of an internal phase [Fig.~\ref{model-fig}]. Importantly, we consider our system as effectively liquid-like, assuming that the phase dynamics are slow with respect to the characteristic timescale of cell rearrangements and the relaxation of elastic stress~\cite{Zehnder2015, THIAGARAJAN2022105053}. Although the underlying particle configuration remains confluent, which may lead to a rigidity transition at high density~\cite{Hannezo-glassy-tissue, PhysRevX.6.021011}, we neglect any solidification here. We show that these minimal assumptions suffice to capture the existence of three main states: a globally synchronized pulsating state, a homogeneous arrested state with constant phase, and a state with propagating radial waves [Fig.~\ref{model-fig1}]. These states are the main defining features of pulsating active liquids, as already reported in some particle-based models~\cite{Togashi2019,Yiwei-Etienne-PAM,Liu_2024,li2024fluidization, pineros2024biased,manacorda2023pulsating}. 

Beyond the obvious connections to continuum mechanics descriptions of tissues mentioned above, our work belongs to the broad class of continuum models of reaction-diffusion systems~\cite{Aranson2002, LINDNER2004321} which encompass oscillatory behaviour~\cite{ALIEV1996293, GANI201630, FN-review}, as well as instabilities of coupled conserved and non-conserved fields~\cite{Sakaguchi-Maeyama, tirtha-oscillator}. Previous works have indeed reported the emergence of pacemakers and travelling waves in excitable media~\cite{Cross_Greenside_2009, hohenberg-review, Sakaguchi-Maeyama}. The uniqueness of our theory, however, lies in its phenomenology of synchronized propagating density and phase waves, as well as spontaneous emergence of a dynamically arrested state.

The paper is organized as follows. In Sec.~\ref{sec:model}, we motivate and present our model of pulsating active liquids. We postulate a hydrodynamic theory from phenomenological and mechanical considerations, report the emergence of propagating waves, and detail the corresponding phenomenology. We discuss the linear stability analysis in Sec.~\ref{sec:linear-stability}, including the role of fluctuations, which leads to analytical predictions for the phase boundaries.  In Sec.~\ref{app:append-coarse-graining}, we argue that our hydrodynamic model embodies the coarse-grained dynamics of a broad class of particle-based models. Finally, we present our main conclusions and future perspectives in Sec.~\ref{sec:summ}.


\section{Pulsating active liquids}\label{sec:model}

Inspired by the phenomenology of pulsatile tissues, we aim to describe the collective behavior of a dense assembly of pulsating particles, where each particle is subject to a periodic drive of an internal degree of freedom which promotes its deformation [Fig.~\ref{model-fig}(a)].  In this section, we propose a minimal, yet non-trivial continuum model capturing the hydrodynamics of such a system, and argue that the emerging nonequilibrium phenomenology can be rationalized in terms of an effective landscape which encapsulates the competition between various steady states.

\subsection{Hydrodynamics of density and phase}\label{sec:fields}

We endow each particle $i$ with an internal degree of freedom $\psi_i$ that represents the phase of an underlying oscillation. For instance, this phase can refer to an internal cellular clock driven by external electrical~\cite{karma-annrev} or chemical signals~\cite{Elad2017}. Assuming that the phase varies slowly in space, as a result of mechanical and biochemical couplings, we define the coarse-grained field $\psi({\bf r},t) $ to capture the spatio-temporal dynamics of the phase at hydrodynamic scales. Similarly, we introduce the local particle density $\rho({\bf r},t)$, which is another field depending on both space and time [Figs.~\ref{model-fig}(b)-(d)].

In the absence of any division and death of particles, the balance of particle numbers is expressed by the conservation equation
\begin{equation}\label{eq-Cont}
    \partial_t \rho = -\nabla \cdot (\rho  {\bf v}) \ ,
\end{equation}
where ${\bf v}$ is the cell velocity field. Force balance implies that external forces compensate for the gradients of the stress tensor $\Sigma$. In the presence of a substrate, we consider that external forces are given by the friction $-\gamma\rho{\bf v}$ and a noise $\sqrt{2D_\rho} \bm\eta_\rho$, yielding
\begin{equation}\label{ext-force-balance}
   0 = - \gamma \rho {\bf v} + \gamma \sqrt{2D_\rho} \bm\eta_\rho + \nabla \Sigma \ ,
\end{equation}
where $\gamma >0$ is an effective friction constant, and $\bm{\eta}_\rho$ has Gaussian statistics with zero-mean and unit variance:
\begin{equation}
    \langle\eta_{\rho,\alpha}({\bf r},t)\eta_{\rho,\beta}({\bf r}',t') \rangle = \delta_{\alpha\beta} \delta({\bf r}-{\bf r}')\delta(t-t') \ .
\end{equation}
For simplicity, we neglect shear stresses and consider only the effect of a finite isotropic compressibility; see Appendix~\ref{viscosity} for model equations with a finite shear viscosity. Specifically, we consider
the following constitutive equation:
\begin{equation}\label{eq:stress}
    \Sigma_{ij} = \lambda \frac{\rho_{\rm ref} -\rho}{\rho_0} \delta_{ij} = - \rho_0 \frac{\delta F}{\delta \rho} \delta_{ij} \ ,
\end{equation}
where $\rho_{\rm ref}$ is a reference density at which the stress vanishes, and $\lambda > 0$ is the compressibility of the tissue. We suppose that the constitutive equation can be obtained from an effective free energy $F$, as expressed by second equality in Eq.~\eqref{eq:stress}. Moreover, we assume that $\rho_{\rm ref}$ changes locally as a function of the phase $\psi$ according to
\begin{equation}\label{ref_elastic_den}
    \rho_{\rm ref}(\psi) = \rho_0 (1+ \epsilon \cos \psi) \ ,
\end{equation}
where $\rho_0$ is the baseline reference density, and $0\le\epsilon<1$ measures the strength of the modulation of the reference density during a cycle. The dependence of $\rho_{\rm ref}$ on $\psi$ captures the change of preferred area of the cells with the internal phase $\psi$; in a stress-free configuration, a change in the reference density will lead to a corresponding change in the cell density $\rho$, as can be seen from Eq.~\eqref{eq:stress}. Taken together, Eqs.~(\ref{eq-Cont}-\ref{eq:stress}) describe how local deviations of the density $\rho$ from $\rho_{\rm ref}(\psi)$ lead to gradients of the stress $\Sigma$, which in turn generates tissue flows.

We now specify the assumptions underlying the dynamics of $\psi$. The pulsation of cells can be driven  by either a cell-autonomous internal clock or a global drive; it favors oscillations of $\psi$ at the same frequency across the whole tissue, yet without imposing a uniform profile of $\psi$ a priori. The feedback between $\rho$ and $\psi$ is described by the coupling term in the 
free energy $F$. Moreover, we assume that interactions between neighboring cells tend to locally synchronize their phases. We capture such an effect with an additional contribution to $F$ describing the energetic cost for stabilizing gradients of $\psi$. Specifically, the free energy reads
\begin{equation}\label{comb-Free-en}
    F[\rho, \psi] = \int_V d{\bf r} \left[ \frac{\lambda}{2} \left( \frac{\rho -\rho_{\rm ref}(\psi)}{\rho_0} \right)^2 + \frac{\kappa}{2} (\nabla \psi)^2 \right] \ ,
\end{equation}
where $\kappa>0$ is the phenomenological parameter that penalizes the formation of interfaces, and $V$ the size of the system. Using Eqs.~\eqref{eq-Cont}-\eqref{eq:stress}, 
the dynamics for $\rho$ and $\psi$ can then be written in a closed form as
\begin{equation}\label{eq:dyn}
\begin{aligned}
    \partial_t \rho &= \nabla \cdot 
    \left( \frac{\rho_0}{\gamma} \nabla \frac{\delta F}{\delta \rho} + \sqrt{2 D_{\rho}} \bm {\eta}_\rho\right)
    \ , 
    \\
    \partial_t \psi  &= \omega - \mu \frac{\delta F}{\delta \psi} + \sqrt{2 D_\psi} \, \eta_{\psi} \ , 
\end{aligned}
\end{equation}
where $\omega>0$ is the driving frequency, and $\mu$ is a kinetic coefficient. The first free-energy term $\delta F/\delta\rho$ simply stems from the compact formulation of the constitutive equation [Eq.~\eqref{eq:stress}] in terms of the free energy $F$. The second free-energy term $\delta F/\delta\psi$ can be regarded as a density-dependent resistance to cycling. The noise term $\eta_\psi$ is uncorrelated with $\bm{\eta}_\rho$, and has Gaussian statistics with zero mean and correlations given by
\begin{equation}
     \langle \eta_{\psi}({\bf r},t)  \eta_{\psi}({\bf r}',t') \rangle = \delta({\bf r}-{\bf r}')\delta(t-t') \ .
\end{equation}
At thermodynamic equilibrium ($\omega=0$), the noise amplitude $D_\rho$ and $D_\psi$ are not independent, since they are determined by the fluctuation-dissipation theorem. In such a case, they must satisfy $D_\rho = {\rho_0 T}/{\gamma}$ and $D_\psi=\mu T$, where $T$ denotes the temperature of the surrounding heat bath. Out of equilibrium ($\omega\neq0$), there is no reason for these constraints to hold a priori.

In short, we postulate here the dynamics of $\rho$ and $\psi$ by combining mechanical and phenomenological arguments at the hydrodynamic level. Importantly, we demonstrate in Sec.~\ref{app:append-coarse-graining} that a similar hydrodynamics can be obtained by coarse-graining the dynamics of pulsating deformable particles. In that respect, Eq.~\eqref{eq:dyn} embodies the hydrodynamics of a broad class of microscopic models, which feature deforming particles with an internal pulsation. In these models, such a pulsation can stem from either (i)~an explicit drive at the microscopic level~\cite{Yiwei-Etienne-PAM, Liu_2024, li2024fluidization, pineros2024biased, manacorda2023pulsating}, or (ii)~from some feedback in the internal state space of particles~\cite{Hannezo-PRXLife, Shiladitya-arxiv23}, as often assumed in excitable media. In both cases, we expect that the coarse-grained behavior will map into a hydrodynamics akin to Eq.~\eqref{eq:dyn}.

In the language of field theories, commonly used to describe the hydrodynamic behavior of soft and active systems~\cite{hohenberg-review, Bray1994, active-matter-marchetti, Cross_Greenside_2009}, the equilibrium limit ($\omega=0$) of Eq.~\eqref{eq:dyn} reduces to a passive model C coupling conserved and non-conserved scalar fields, respectively $\rho$ and $\psi$: the corresponding equilibrium model cannot accommodate any steady current. In contrast, as we discuss in more detail in the following sections, the presence of pulsation ($\omega>0$) opens the door to a rich nonequilibrium behavior, with the emergence of propagating waves [Fig.~\ref{model-fig1}] reminiscent of the phenomenology of pulsating tissues~\cite{karma-annrev,collinet2021programmed}.


\subsection{Effective landscape and irreversibility}
\label{sec:land}

In what follows, we consider a non-dimensional version of the dynamics in two spatial dimensions. To this end, we scale time and space variables as
\begin{equation}
    t \rightarrow t' = t/t_{\rm c} \ ,
    \quad
    {\bf r} \rightarrow {\bf r}' = {\bf r} / \ell_{\rm c} \ ,
\end{equation}
in terms of the characteristic time and length scales:
\begin{equation}\label{eq:scale}
    t_{\rm c}=1/\omega \ ,
    \quad
    \ell_{\rm c} = \sqrt{\mu\kappa/\omega} \ .
\end{equation}    
We also scale density variables and noise amplitudes as
\begin{equation}
\begin{aligned}
    \rho &\rightarrow \rho' =\rho/\rho_0 \ ,
    \quad\qquad
    \rho_{\rm ref} \rightarrow \rho_{\rm ref}'=\rho_{\rm ref}/\rho_0 \ ,
    \\
    D_\rho &\rightarrow D'_\rho = \frac{D_\rho\omega}{(\mu\kappa\rho_0)^2} \ ,
    \quad
    D_\psi \rightarrow D'_\psi = \frac{D_\psi}{\mu\kappa} \ .
\end{aligned}
\end{equation}
The non-dimensional dynamics can then be written in terms of the free-energy density
\begin{equation}\label{eq:f}
    f(\rho,\psi) = (\beta /2) ( 1 + \epsilon \cos \psi -\rho)^2
\end{equation}
as
\begin{equation}\label{rho-modelB-noise}
\begin{aligned}
    \partial_{t} \rho &= \frac{\alpha}{\beta} \nabla^2 \frac{\partial f}{\partial\rho} + \sqrt{2 D_{\rho}} \nabla  \cdot\bm{\eta}_{\rho} \ ,
    \\
    \partial_{t} \psi &= \nabla^2 \psi - \frac{\partial}{\partial\psi} (f-\psi) + \sqrt{2 D_{\psi}} \, \eta_{\psi} \ ,
\end{aligned}
\end{equation}
where we have omitted the prime notation for simplicity, and introduced the non-dimensional parameters
\begin{equation}
    \alpha = \lambda/(\mu\kappa\gamma \rho_0) \ , 
    \quad
    \beta = \mu \lambda/\omega \ .
\end{equation}
The total density $\bar\rho=\frac{1}{V}\int_V d{\bf r} \rho$ is conserved and constant for a given system size $V$. In the noiseless dynamics ($D_\rho=0$ and $D_\psi=0$), we are thus left with four independent control parameters $(\bar\rho, \alpha, \epsilon, \beta)$ which are all positive. Given that $f-\psi$ is unbounded, it should not be regarded as an equilibrium free energy. Yet, as discussed in the next section, analyzing the features of $f-\psi$ provides some insights into the various types of emerging states. We also discuss in Appendix~\ref{sec:current_closed} how our dynamics can be written in a closed form in terms of the density $\rho$ and the magnetization ${\bf m} = \epsilon a \nabla ( \cos \psi)$, where  $a = \lambda/(\gamma \rho_0)$, thus providing an alternative perspective on our description of pulsating liquids.

To assess the deviation from equilibrium of our model [Eq.~\eqref{rho-modelB-noise}], we want to quantify the irreversibility of the corresponding trajectories of $[\rho,\psi]_0^\tau$, where $\tau$ refers to the trajectory length~\cite{Jack2022}. To this end, we introduce the path probability of such trajectories ${\cal P}\sim e^{-\cal A}$ in terms of the dynamic action ${\cal A}={\cal A}_\rho+{\cal A}_\psi$, where
\begin{equation}\label{eq:action}
\begin{aligned}
    {\cal A}_\rho &= \frac{1}{4D_\rho} \int_0^\tau dt \int_V d{\bf r} \left[ \nabla^{-1} \left(\partial_t\rho - \frac{\alpha}{\beta} \nabla^2 \frac{\partial f}{\partial\rho} \right) \right]^2 \ ,
    \\
    {\cal A}_\psi &= \frac{1}{4D_\psi} \int_0^\tau dt \int_V d{\bf r} \left( \partial_t\psi - 1 - \nabla^2\psi + \frac{\partial f}{\partial\psi} \right )^2 \ .
\end{aligned}
\end{equation}
In all what follows, we use Stratonovich convention, for which the actions in Eq.~\eqref{eq:action} should feature some extra terms~\cite{Nardini2017}. In fact, these terms are invariant under time reversal, so that we can safely omit them for our purpose. We consider the Kullback-Leibler divergence ${\cal D}_{\rm KL}$ between the forward and time-reversed dynamics, with path probabilities respectively denoted $({\cal P}, {\cal P}_{\rm R})$, as
\begin{equation}
    {\cal D}_{\rm KL} = \underset{\tau\to\infty}{\lim} \frac{1}{\tau} \ln \frac{\cal P}{{\cal P}_{\rm R}} \ ,
\end{equation}
where the time-reversal operation amounts to substituting $(\partial_t\rho,\partial_t\psi)\to -(\partial_t\rho,\partial_t\psi)$ into Eq.~\eqref{eq:action}, yielding
\begin{equation}
	{\cal D}_{\rm KL} = \underset{\tau\to\infty}{\lim} \frac{1}{\tau} \big( {\cal A}_{\rho,\rm R} - {\cal A}_{\rho} + {\cal A}_{\psi,\rm R} - {\cal A}_{\rho} \big) \ ,
\end{equation}
where
\begin{equation}
\begin{aligned}
    {\cal A}_{\rho,\rm R} - {\cal A}_{\rho} &= - \frac{\alpha}{\beta D_\rho}\int_0^\tau dt \int_V d{\bf r} ~ \frac{\partial f}{\partial\rho} \partial_t \rho \ ,
    \\
    {\cal A}_{\psi,\rm R} - {\cal A}_{\psi} &= - \frac{1}{D_\psi}\int_0^\tau dt \int_V d{\bf r} ~ \bigg( \frac{\partial f}{\partial\psi} - 1 - \nabla^2 \psi \bigg) \partial_t\psi \ .
\end{aligned}
\end{equation}
When mobilities and noise amplitudes are related as in equilibrium ($\alpha D_\psi = \beta D_\rho$), we deduce
\begin{equation}\label{eq:epr}
\begin{aligned}
    {\cal D}_{\rm KL} &= \underset{\tau\to\infty}{\lim} \frac{{\cal F}(0) - {\cal F}(\tau)}{D_\psi \tau} + \frac{V}{D_\psi} \nu ,
    \\
    {\cal F} &= \int_V d{\bf r} \bigg[ f(\rho,\psi) + \frac{1}{2} (\nabla\psi)^2 \bigg] \ ,
\end{aligned}
\end{equation}
with the chain rule $\dot {\cal F} = \int_V d{\bf r} \big[ (\delta {\cal F}/\delta \rho)\partial_t\rho + (\delta {\cal F}/\delta\psi) \partial_t\psi \big]$ being valid for Stratonovich convention~\cite{Gardiner}.  The current $\nu$ reads
\begin{equation}\label{eq:nu}
    \nu = \underset{\tau\to\infty}{\lim} \frac{1}{V \tau} \int_0^\tau dt \int_V d{\bf r} \ \partial_t \psi = \frac 1 V \int_V d{\bf r} \langle \partial_t \psi \rangle \ ,
\end{equation}
where the average $\langle\cdot\rangle$ is performed over time and realizations, under the ergodic assumption. Since $f$ is bounded [Eq.~\eqref{eq:f}], we deduce that the contribution of the free energy $\cal F$ to ${\cal D}_{\rm KL}$, given by the first term in Eq.~\eqref{eq:epr}, vanishes. The additional contribution featuring $\nu>0$, that is not present in passive model C~\cite{Chaikin}, yields ${\cal D}_{\rm KL}>0$. Therefore, the presence of phase current is a direct signature of the irreversibility of the dynamics, as befits any nonequilibrium model.


\subsection{Competition between cycles and arrest}\label{sec:comp}

We now discuss the phenomenology of the homogeneous states emerging from Eq.~\eqref{rho-modelB-noise} in the absence of noise ($D_\rho=0$ and $D_\psi=0$). If one starts from a homogeneous initial condition, the dynamics is entirely determined by the evolution of the homogeneous phase $\psi$, since the density $\rho=\bar\rho$ remains constant at all times. Therefore, the dynamics of the homogeneous states is not affected by $\alpha$, and depends only on $(\bar\rho, \epsilon, \beta)$.

\begin{figure}
    \includegraphics[width=0.99\columnwidth]{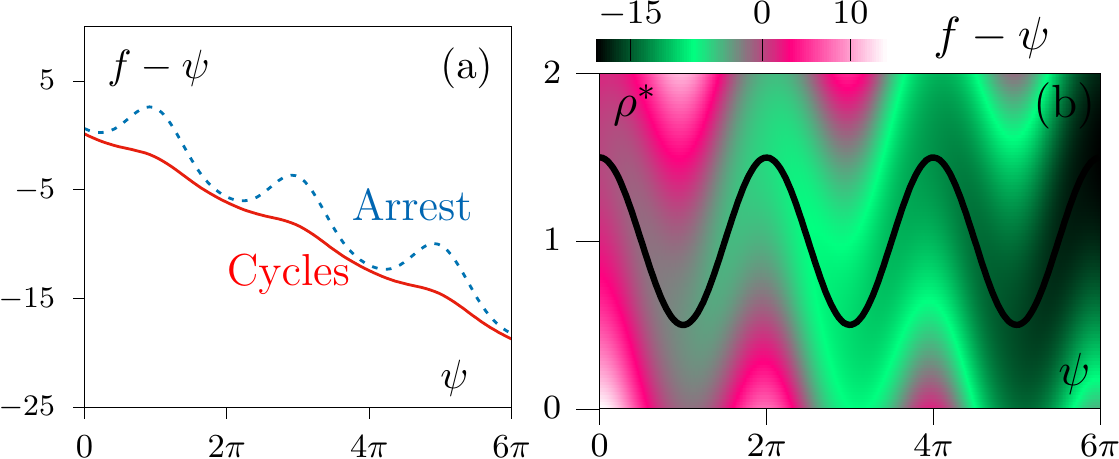}
    \caption{(a,b)~Landscape $f(\rho,\psi)-\psi$. For fixed $\rho$, the phase $\psi$ arrests at low drive ($\beta>\beta_{\rm c}(\bar\rho)$, dotted blue), and cycles at high drive ($\beta<\beta_{\rm c}(\bar\rho)$, solid red). The black line in (b) is $\rho^*(\psi) = {\rm argmin}_\rho f(\rho,\psi)$. Parameters: $\epsilon=0.5$, (b)~$\beta=12.5$.
    }
\label{landscape}
\end{figure}

The landscape $f(\rho,\psi)-\psi$ features a series of minima if the inverse drive $\beta$ is higher than a critical value $\beta_{\rm c}$ at fixed density $\rho$ [Fig.~\ref{landscape}(a)]. Therefore, the parameters $(\bar\rho,\beta)$ naturally control the transition between two distinct states: (i)~a cycling state with a steady current, for which $\beta<\beta_{\rm c}(\bar\rho)$, and (ii)~an arrested state without any steady current, for which $\beta>\beta_{\rm c}(\bar\rho)$. The competition between arrest and cycles is an essential feature  of
pulsating active matter~\cite{Yiwei-Etienne-PAM, manacorda2023pulsating,  Liu_2024, li2024fluidization, pineros2024biased}. Specifically, the existence of arrest stems from the breakdown of invariance with respect to a phase shift: in contrast with the standard CGLE~\cite{aranson-review}, our dynamics [Eq.~\eqref{rho-modelB-noise}] is not invariant under a phase shift $\psi \to \psi + C$ for an arbitrary constant $C$. Therefore, our hydrodynamic model clearly reproduces the unique phenomenology of pulsating liquids at homogeneous level, as already reported in particle-based models~\cite{Togashi2019, Yiwei-Etienne-PAM, manacorda2023pulsating, Liu_2024, li2024fluidization, pineros2024biased}.

In the arrested state, the stationary solution $\psi=\bar\psi$ is a local minimum of the landscape $f(\rho, \psi) - \psi$. Therefore, such a solution obeys
\begin{equation}\label{eq:arr}
    \frac{\partial f}{\partial\psi}(\bar\psi) = 1 \ ,
    \quad
    \frac{\partial^2 f}{\partial\psi^2}(\bar\psi) > 0 \ ,
\end{equation}
namely
\begin{equation}\label{hom-psistar-evol}
\begin{aligned}
    0 &= 1+ \beta \epsilon (1 - \bar\rho + \epsilon  \cos \bar\psi)\sin \bar\psi \ ,
    \\
    0 &< (1-\bar\rho) \cos \bar\psi + \epsilon \cos 2\bar\psi \ .
\end{aligned}
\end{equation}
The phase boundary $\beta_{\rm c}(\bar\rho)$ corresponds to the case where the local minimum and the inflection point of $f$ coincide. The corresponding solution $\psi=\psi_c$ satisfies the following condition:
\begin{equation}
    \frac{\partial f}{\partial\psi}(\psi_{\rm c}) = 1 \ ,
    \quad
    \frac{\partial^2 f}{\partial\psi^2}(\psi_{\rm c}) = 0 \ ,
    \quad
    \frac{\partial^3 f}{\partial\psi^3}(\psi_{\rm c}) < 0 \ ,
\end{equation}
namely
\begin{equation}
\begin{aligned}
    0 &= 1+ \beta_{\rm c} \epsilon (1 - \bar\rho + \epsilon  \cos \psi_{\rm c})\sin \psi_{\rm c} \ ,
    \\
    0 &= (1-\bar\rho) \cos \psi_{\rm c} + \epsilon \cos 2\psi_{\rm c} \ ,
    \\
    0 &> (1-\bar\rho) \sin \psi_{\rm c} + 2 \epsilon \sin 2\psi_{\rm c} \ .
\end{aligned}
\end{equation}
leading to
\begin{equation}\label{homogeneous-phase-boundary1}
    \beta_{\rm c}(\bar\rho) = \frac{1}{\epsilon (\bar\rho-1- \epsilon \cos \psi_{\rm c} ) \sin\psi_{\rm c}} \ ,
\end{equation}
where
\begin{equation}
\begin{aligned}
    \cos \psi_{\rm c} &= \frac{1}{4}\left[ (\bar\rho-1)/\epsilon - \sqrt{ ((\bar\rho-1)/\epsilon)^2+8} \right]  \, \text{if} \; \bar\rho\geq 1 \ ,
    \\
    \cos \psi_{\rm c} &= \frac{1}{4}\left[ (\bar\rho-1)/\epsilon + \sqrt{ ((\bar\rho-1)/\epsilon)^2+8} \right]  \, \text{if} \; \bar\rho<1 \ .
\end{aligned}
\end{equation}
Therefore, Eq.~\eqref{homogeneous-phase-boundary1} shows that our simple, yet non-trivial model of pulsating active liquids entails an exact analytical prediction for the phase boundary $\beta_{\rm c}(\bar\rho)$ delineating the existence of either arrested or cycling states.

Interestingly, analyzing further the shape of $f$ in the space $(\rho,\psi)$ allows one to anticipate the behavior of the system even beyond the case of its homogeneous states. Indeed, at each position ${\bf r}$, the system explores the landscape both in terms of $\rho$ and $\psi$. As the system evolves towards increasing $\psi$, the local minimum of the landscape $f$ corresponds to oscillating $\rho$ [Fig.~\ref{landscape}(b)]: this mechanism suggests that the cycling of phase favors sustained oscillations of density. Since the density is conserved, any local decrease in density must be compensated for by a corresponding increase elsewhere in the system. Therefore, oscillations of density can only create gradients, yielding the formation of oscillating patterns, as we discuss in more detail in the next section.


\begin{figure*}
    \includegraphics[width=2\columnwidth]{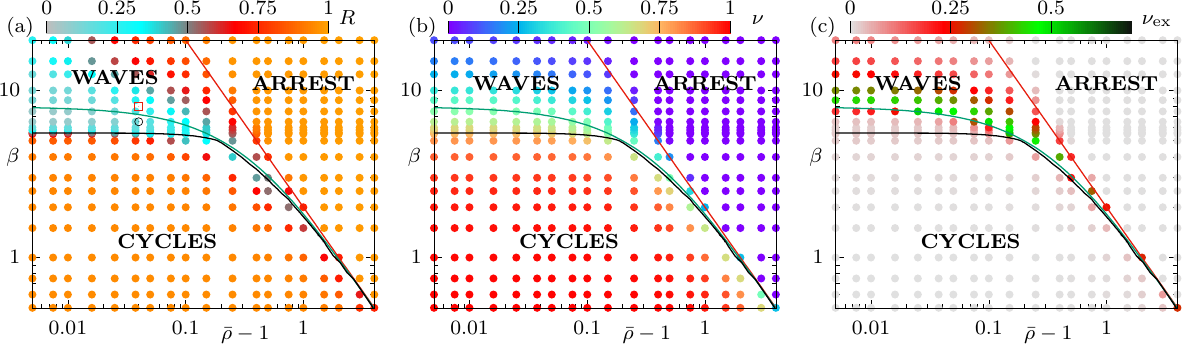}
    \caption{Phase diagram in terms of the inverse drive $\beta$ and the total density $\bar\rho =\frac 1 V \int d{\bf r} \rho$.
    (a)~The synchronization parameter $R$ [Eq.~\eqref{eq:R}], (b)~the phase current $\nu$ [Eq.~\eqref{eq:R}], and (c)~the excess current $\nu_{\rm ex}$ [Eq.~\eqref{nu-net_defn}] allow one to distinguish three regimes associated with distinct states. Arrest emerges at large $\beta$ and large $\bar\rho$, cycles at small $\beta$ and $\bar\rho\simeq 1$, whereas patterns appear for intermediate regimes between arrest and cycles. The green line denotes the boundary of existence between arrested and cycling states [Eq.~\eqref{homogeneous-phase-boundary1}]. The red and black lines delineate the boundary of instability for arrest [Eq.~\eqref{pert-pb}] and cycles [Sec.~\ref{sec:floquet}], respectively.
    Parameters: $\epsilon=0.5$, $\alpha=1$, $D_\rho= D_\psi=0.1$. The hollow symbols (square and circle) in panel (a) refer to the parameter values used in Fig.~\ref{pattern-2-fig}. Appendix~\ref{app:num} gives details on our numerical methods.}
\label{phase_diagram}
\end{figure*}

\subsection{Synchronization and current}

Internal pulsation leads to patterns of density and
phase propagating throughout the system. To identify
the regimes where such patterns emerge, we introduce
three order parameters quantifying the overall synchronization and the current of phase. We report the corresponding phase diagram [Fig.~\ref{phase_diagram}], and discuss the phenomenology of the patterns observed numerically by simulating Eqs.~\eqref{rho-modelB-noise}.

We quantify synchronization $R$ and phase current $\nu$:
\begin{equation}\label{eq:R}
    R = \bigg\langle \bigg| \int_V e^{i \psi} \frac{d{\bf r}}{V} \bigg| \bigg\rangle ,
    \quad
    \nu = \int_V \frac{\langle \partial_t \psi \big\rangle}{\omega} \frac{d{\bf r}}{V} ,
\end{equation}
where $\langle\cdot\rangle$ indicate the average over time and realizations. Full synchronization ($R=1$) corresponds to homogeneous configurations (either cycles or arrest), and incomplete synchronization ($R<1$) points to pattern formation. Homogeneous arrest ($R\simeq1$ and $\nu\simeq0$) and cycles ($R\simeq1$ and $\nu > 0$) are, respectively, stable for large and small $(\beta,|\bar\rho-1|)$, as expected [Figs.~\ref{phase_diagram}(a,b)]. When incomplete synchronization emerges ($R<1$), the current deviates from its limit values ($0<\nu<1$).

To quantify the degree of inhomogeneity in the profiles $(\rho, \psi)$, we compare the current $\nu$ with its value for the deterministic, homogeneous system by introducing the excess current $\nu_{\rm ex}$ as
\begin{equation}\label{nu-net_defn}
 \nu_{\rm ex} = \nu - \langle \dot \Psi\rangle/\omega ,
\end{equation}
where $\Psi$ is the homogeneous, noiseless realization of the phase: $\dot\Psi = 1 -  (\partial f/\partial\Psi)(\bar\rho,\Psi)$. Arrest and cycles [Fig.~\ref{landscape}] are associated with $\nu_{\rm ex}\simeq0$ [Fig.~\ref{phase_diagram}(c)]. In contrast, waves lead to $\nu_{\rm ex}>0$, revealing that the inhomogeneous phase $\psi$ cycles faster than its homogeneous counterpart $\Psi$. In fact, $\nu_{\rm ex}$ is anticorrelated with $R$, so the highest $\nu_{\rm ex}$ coincides with the lowest $R$, specifically at $\omega\simeq \omega_{\rm c}(\bar\rho)$. These results corroborate that $(R,\nu)$ and $\nu_{\rm ex}$ consistently detect the emergence of waves.



\begin{figure*}
    \includegraphics[width=1.8\columnwidth]{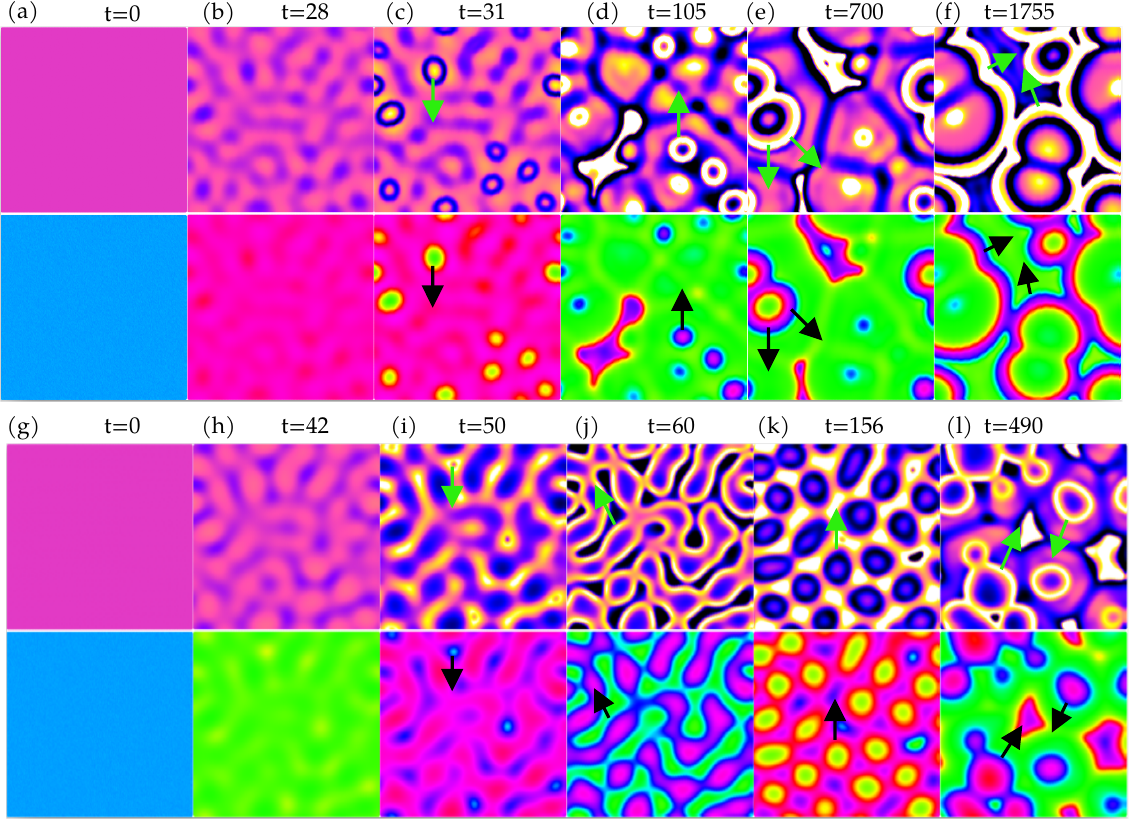}
    \caption{Time evolution of the density $\rho$ [top panel] and the phase $\psi$ [bottom panel] fields in the regimes of (a-f)~unstable arrest ($\beta_{\rm c}<\beta=9<\beta_{\rm ar}$ and (g-l)~unstable cycles ($\beta_{\rm cy}<\beta=6.5<\beta_{\rm c}$, without any noise ($D_\rho=0$ and $D_\psi=0$); see~\cite{Movie2} for a movie. Starting from a homogeneous stationary profile of $(\rho,\psi)$, a spinodal instability first leads to the formation of some growing domains. At a later stage, when the local density goes below a critical value, waves emerge and propagate in the system. Arrows mark directions of propagation. Other parameters : $\bar\rho=1.04$, $\epsilon=0.5$,  $\alpha=1$, $dx=0.25$, $dt=0.001$ and $V=64 \times 64$. Colormaps as in Fig.~\ref{model-fig}.
    }
\label{pattern-2-fig}
\end{figure*}

\subsection{Propagating waves}\label{sec:waves}

In the regimes where both uniform profiles are unstable, initial perturbations lead to coarsening domains of high and low densities [Fig.~\ref{pattern-2-fig}]. Such domains grow in a background which is either arrested ($\beta>\beta_{\rm c}$) or cycling ($\beta<\beta_{\rm c}$). In both cases, domains with lower densities encounter another instability, reminiscent of secondary bifurcations~\cite{Sakaguchi-PTP93, Coullet-Iooss-PRL90}, yielding the emergence of propagating radial waves. In contrast to some recent studies of the  hydrodynamics of pulsating particles~\cite{Yiwei-Etienne-PAM, manacorda2023pulsating}, the waves propagate in the profile of both density and phase. Indeed, $\rho$ cannot remain homogeneous in the presence of gradients in $\psi$ [Eq.~\eqref{rho-modelB-noise}]. Interestingly, the formation of such waves can be rationalized in terms of the local effective free-energy picture [Eq.~\eqref{eq:f}].

In the case of an arrested background [Figs.~\ref{pattern-2-fig}(a-f)], the local phase escapes from its background value at specific locations where the local density goes below a given threshold: consistently with the effective landscape picture [Sec.~\ref{sec:land}], reducing the local density is indeed a route to promoting a cycling phase. The cycling phase spontaneously organizes into propagating waves. In the case of a cycling background [Figs.~\ref{pattern-2-fig}(g-l)], a non-uniform density leads to inhomogeneities in the cycling frequency, which in turn induces a lag between the phases of nearby domains. This phase lag leads to fronts propagating throughout the system.

Unstable domains turn into regions where the density oscillates, which we refer to as {\it pacemakers}~\cite{Cross_Greenside_2009}, from which radial waves emanate and propagate across the system. For large enough systems, multiple pacemakers coexist simultaneously.  Eventually, waves spontaneously organize into temporally oscillating domains [Fig.~\ref{model-fig1}]. The relative diffusion of the coupled fields, and therefore the speed of the propagating waves, is controlled by $\alpha$ [Eq.~\eqref{rho-modelB-noise}]. Specifically, waves move slower (faster) for smaller (larger) values of $\alpha$. Therefore, for small $\alpha$, propagating waves leave behind a locally and transiently homogeneous region. The pacemakers are at the center of such regions, and the boundaries between the regions correspond to the annihilation loci of the waves. In the companion short paper, we show that the spatio-temporal organization of pacemakers and travelling waves can interestingly be understood in terms of the dynamics of specific topological defects in the orientation of phase gradients~\cite{ASP}.

Investigating the patterns in the fluctuation-destabilized zone in the phase diagram, we observe that the instabilities become propagating pulses rather than waves. Fluctuations lead to nucleation events where an unstable domain coarsens to a certain size, and then undergoes a secondary instability to eject a contraction pulse. 



\section{Stability analysis}\label{sec:linear-stability}

The mechanisms destabilizing the homogeneous state, either arrested [Figs.~\ref{pattern-2-fig}(a-f)] or cycling [Figs.~\ref{pattern-2-fig}(g-l)], can be examined using linear analysis. We first show that the arrested state undergoes a spinodal instability with coarsening of the density and phase profiles. Then, we explore how fluctuations modify the effective landscape, and thus affect the stability of arrest. Finally, we investigate how cycling states get destabilized using a Floquet analysis.

\subsection{Spinodal instability of arrest}\label{sec:stab-arr}

To carry out a linear stability analysis of the arrested phase, we expand $(\rho,\psi)$ around their respective homogeneous solutions $(\bar\rho,\bar\psi)$, where $\bar\rho=\frac 1 V \int_V d{\bf r} \rho$ denotes the total density, and $\bar\psi$ obeys Eq.~\eqref{hom-psistar-evol}. The perturbations in Fourier space are then given by
\begin{equation}
\begin{aligned}
    \rho_q &= \frac 1 V \int_V d{\bf r} e^{i {\bf q}\cdot{\bf r} } \big[ \rho({\bf r},t) - \bar\rho\big] \ ,
    \\
    \psi_q &= \frac 1 V \int_V d{\bf r} e^{i {\bf q}\cdot{\bf r} } \big[ \psi({\bf r},t) - \bar\psi\big] \ .
\end{aligned}
\end{equation}
The corresponding linearized dynamics can be written as
\begin{equation}
\label{eq:lin_dynamics_arrested}
\frac{d}{dt}
\begin{bmatrix}
    \rho_q \\ \psi_q
\end{bmatrix}
= {\mathbb S}_{\rm A}(q) 
\begin{bmatrix}
    \rho_q \\ \psi_q
\end{bmatrix} .
\end{equation}
The stability matrix ${\mathbb S}_{\rm A}$ depends on $q=|{\bf q}|$ as
\begin{equation} \label{M0-arrest-dimless}
    {\mathbb S}_{\rm A}(q) =
    \begin{bmatrix}
        -\alpha q^2 & -\alpha a q^2
        \\
        -\beta a &  -  q^2 + e
    \end{bmatrix}
    \ ,
\end{equation}
and is given in terms of
\begin{equation}\label{e-equn}
    a = \epsilon \sin \bar\psi \ ,
    \quad
    e = - \cot\bar\psi - \beta \epsilon^2 \sin^2 \bar\psi < 0 ,
\end{equation}
where we have used Eq.~\eqref{hom-psistar-evol} to simplify the expression of $e$. Analyzing the eigenvalues of ${\mathbb S}_{\rm A}$ provides information about any potential $q-$dependent linear instability~\cite{hohenberg-review, Bray1994}. The eigenvalues $\Lambda_{\pm}$ are of the form
\begin{equation}\label{eq:lambda_p}
    \Lambda_{\pm}(q)=\frac{1}{2}\left( A(q) \pm \sqrt{A^2(q) - 4 q^2 B(q) }\right) \ ,
\end{equation}
where 
\begin{equation}
    A(q) =  e- (\alpha+1)q^2 \ ,
    \quad
    B(q) = \alpha ( q^2 - \beta a^2 - e) \ .
\end{equation} 
Note that $A(q)$ and $q^2 B(q)$ correspond to the trace and determinant of the stability matrix ${\mathbb S}_{\rm A}$, respectively, which are reminiscent of the trace and determinant of the stability matrix that occurs in hydrodynamic theories of active chemorepulsive colloids~\cite{Liebchen-Mike-PRL}.

An instability arises if at least one of the eigenvalues $\Lambda_{\pm}$ has a positive real part. To predict the emergence of such an instability, it suffices to analyze the behavior of $\Lambda_{\pm}$ at small wavenumber $q$, as
\begin{equation}
\begin{aligned}
    \Lambda_+(q) &= - \alpha ( 1 + \beta a^2 / e) q^2 + O(q^4) \ ,
    \\
    \Lambda_-(q) &= e + (\alpha\beta a^2/e - 1) q^2 + O(q^4) \ .
\end{aligned}
\end{equation}
We deduce that $\Lambda_{-}(q)$  is negative for all $q$, whereas $\Lambda_{+}(q)$ can have a positive real part for some $q$ if $\beta a^2 + e<0$. Using the expression of $(a,e)$ [Eq.~\eqref{e-equn}], it follows that the limit of stability ($\beta a^2 + e=0$) corresponds to
\begin{equation}\label{pb-condn-arrest-q}
    \beta \epsilon \cos \bar\psi \left( 1+ \epsilon \cos \bar\psi -\bar\rho \right) = 0 \ .
\end{equation}
The condition in Eq.~\eqref{pb-condn-arrest-q} is equivalent to enforcing that the determinant of ${\mathbb S}_{\rm A}$ is positive, namely $B(q)>0$ for finite $q>0$. Note that this condition is not affected by $\alpha$, given that $\bar\psi$ does not depend on $\alpha$ [Eq.~\eqref{hom-psistar-evol}]. Substituting Eq.~\eqref{hom-psistar-evol} in Eq.~\eqref{pb-condn-arrest-q} yields $\cos\bar\psi=0$, or equivalently $\sin\bar\psi=\pm 1$. Finally, substituting this criterion into Eq.~\eqref{pb-condn-arrest-q} leads to the phase boundary $\beta_{\rm ar}(\bar\rho)$ given by
\begin{equation}\label{pert-pb}
    \beta_{\rm ar}(\bar\rho) = \frac{1}{\epsilon|1-\bar\rho| } .
\end{equation}
As a result, we predict that the part of the phase space $(\beta,\bar\rho)$ contained between $\beta_{\rm c}(\bar\rho)$ [Eq.~\eqref{homogeneous-phase-boundary1}, green line in Fig.~\ref{phase_diagram}] and $\beta_{\rm ar}(\bar\rho)$ [Eq.~\eqref{pert-pb}, red line in Fig.~\ref{phase_diagram}] accommodates a linear instability of the arrested state. Our analytical prediction agrees well with numerical simulations [Fig.~\ref{phase_diagram}]: in the regime of instability ($\beta_{\rm c}<\beta<\beta_{\rm ar}$), $R<1$ [Eq.~\eqref{eq:R}] and $\nu_{\rm net}>0$ [Eq.~\eqref{nu-net_defn}] clearly indicate that the system deviates from the arrested state. Note that both $\beta_{\rm c}$ and $\beta_{\rm ar}$ are independent of $\alpha$.

\begin{figure}
    \includegraphics[width=\columnwidth]{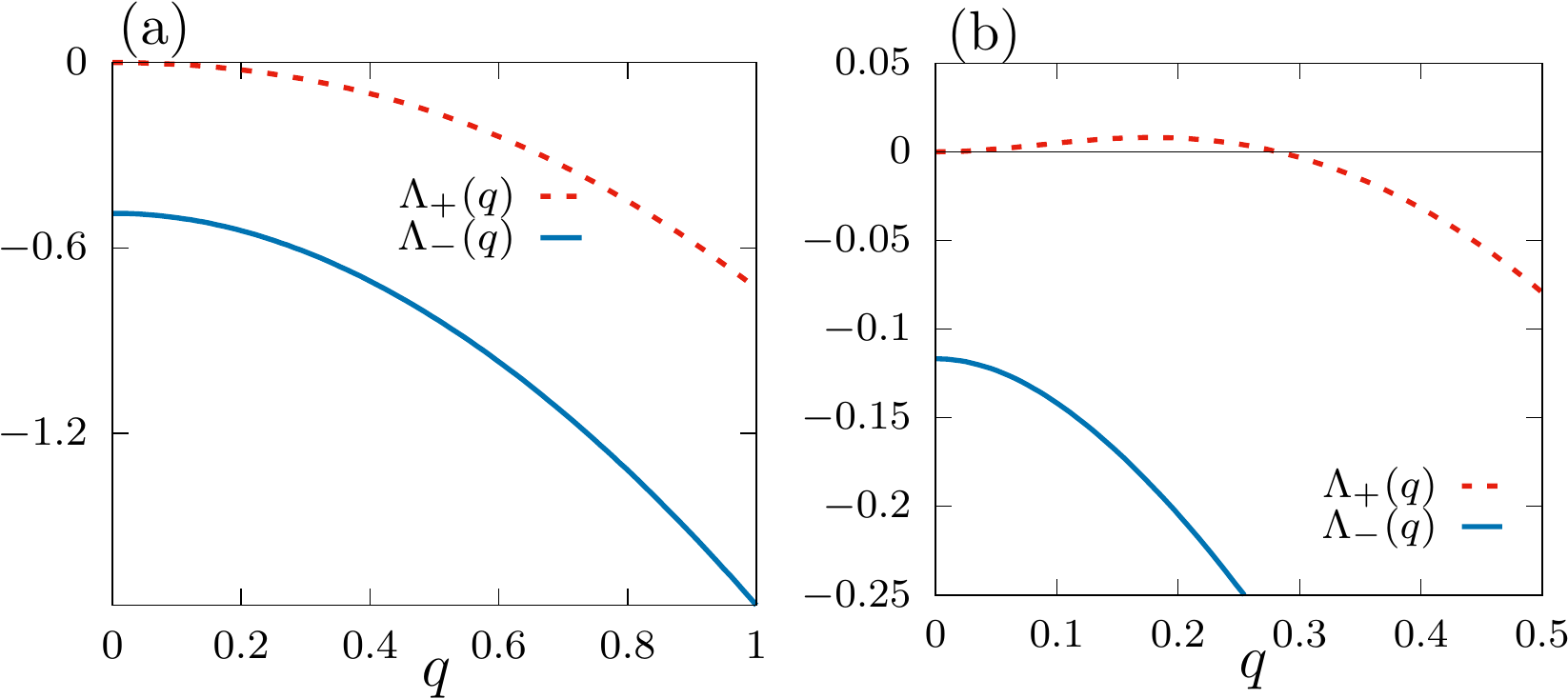}
    \caption{The eigenvalues $\Lambda_{\pm}(q)$ [Eq.~\eqref{eq:lambda_p}] of the stability matrix for the arrested state reveal that the dynamics is either (a)~stable or (b)~unstable with spinodal-like scenario. Parameters: $\epsilon=0.5$, $\alpha=1$, $\bar\rho=3.5$, (a)~$\beta=0.88$, and
    (b)~$\beta=0.79$.
    }
\label{Lam_pm}
\end{figure}

In practice, the growth of perturbations is described by the eigenvalue $\Lambda_+(q)$. Interestingly, the behavior of $\Lambda_+(q)$ is analogous to the case of spinodal instabilities [Fig.~\ref{Lam_pm}], as observed for instance in liquid-liquid phase separation~\cite{hohenberg-review, Bray1994}. For certain choices of model parameters, it is characterized by a finite range of unstable modes from $q=0$ to $q>0$, with the fastest growing mode given by the maximum of $\Lambda_+(q)$. The mode $q=0$ is always stable ($\Lambda_+(0)=0$) since the total density $\bar\rho$ does not evolve in time, so that $\rho_{q=0}=0$. Snapshots in Fig.~\ref{pattern-2-fig}(a-f) show the time evolution of an unstable arrested state.


\subsection{The role of fluctuations}\label{sec:noise-calculations}

It is well-known that fluctuations, driven in our case by Gaussian noise terms [Eq.~\eqref{rho-modelB-noise}], can strongly affect the phase behaviors of excitable systems~\cite{LINDNER2004321}. We now study whether the homogeneous arrested state can be potentially destabilized by the noise terms in the dynamics [Eq.~\eqref{rho-modelB-noise}]. To this end, we are interested in obtaining a closed dynamics for the profiles averaged over noise realizations, which we denote by $\rho_0=\langle\rho\rangle$ and $\psi_0=\langle\psi\rangle$, in the regime of small fluctuations ($D_\rho\ll1$ and $D_\psi\ll1$).

\subsubsection{Dynamics at weak noise}

We follow a strategy analogous to that in Ref.~\cite{Martin-et-al} by decomposing the fluctuating fields as
\begin{equation}
    \rho({\bf r},t) = \rho_0({\bf r},t) + \delta \rho({\bf r},t) \ , 
    \quad
    \psi({\bf r},t) = \psi_0({\bf r},t) + \delta \psi({\bf r},t) \ ,
\end{equation}
and assuming that $(\delta\rho,\delta\psi)$ is a weak correction with respect to $(\rho_0, \psi_0)$. Taking the average of Eq.~\eqref{rho-modelB-noise}, we get
\begin{equation}
    \partial_t \rho_0 = \frac{\alpha}{\beta} \nabla^2 \bigg\langle\frac{\partial f}{\partial\rho} \bigg\rangle \ ,
    \quad
    \partial_t \psi_0 = 1 + \nabla^2 \psi_0 - \bigg\langle \frac{\partial f}{\partial\psi} \bigg\rangle \ ,
\end{equation}
where the landscape $f$ is defined in Eq.~\eqref{eq:f}. We expand the relevant terms in the dynamics up to the second order in the perturbation $(\delta \rho, \delta \psi)$. This leads to the following evolution equations for the fields $(\rho_0, \psi_0)$:
\begin{equation}\label{barrho-0}
\begin{aligned}
    \partial_t \rho_0 &= \alpha \nabla^2 \rho_0 - \alpha \epsilon \nabla^2 \left[ \left(1- \frac{\langle \delta \psi^2 \rangle}{2}\right) \cos \psi_0\right] \ ,
    \\
    \partial_t \psi_0 &= 1+ \nabla^2 \psi_0 + \beta \epsilon \left( 1- \frac{\langle \delta \psi^2\rangle}{2}\right) (1- \rho_0) \sin \psi_0
    \\ 
    &\quad - \beta \epsilon \langle \delta \rho \delta \psi \rangle \cos \psi_0 + \frac{\beta \epsilon^2}{2} (1- 2 \langle \delta \psi^2 \rangle) \sin 2\psi_0 \ .
\end{aligned}
\end{equation}
We have used that $\delta\rho$ and $\delta\psi$ both vanish on average by definition. The dynamics in Eq.~\eqref{barrho-0} does not feature any noise term. Yet, it depends on the noise amplitudes $(D_\rho, D_\psi)$ through the correlators $\langle\delta\psi^2\rangle$ and $\langle\delta\rho\delta\psi\rangle$. In that respect, Eq.~\eqref{barrho-0} embodies how fluctuations affect the dynamics in the regime of weak noise.

The dynamics of the weak perturbation $(\delta\rho, \delta\psi)$ is obtained by linearizing the fluctuating dynamics [Eq.~\eqref{rho-modelB-noise}]. In doing so, inspired by~\cite{Martin-et-al, martin2024}, we assume that $(\delta\rho, \delta\psi)$ relax faster than the fields $(\rho_0, \psi_0)$. Hence, we consider that $(\rho_0, \psi_0)$ are constant, both in space and time, in the dynamics $(\delta\rho, \delta\psi)$. It follows that this dynamics can be written in Fourier space as
\begin{equation}\label{lin-stab-matrix-noise}
    \frac{d}{dt}
    \begin{bmatrix}
    \delta \rho_q \\ \delta \psi_q
    \end{bmatrix}
    =
    {\mathbb L}(q)
    \begin{bmatrix}
    \delta \rho_q \\ \delta \psi_q
    \end{bmatrix} +
    \begin{bmatrix}
    -  i \sqrt{2 D_{\rho}}{\bf q} \cdot \bm{\eta}_{\rho,q} \\ \sqrt{2 D_{\psi}}\eta_{\psi,q}
    \end{bmatrix} ,
\end{equation}
where
\begin{equation}
    {\mathbb L}(q) =
    \begin{bmatrix}
        -\alpha q^2 & -\alpha a_0 q^2
        \\
        -\beta a_0 &  -  q^2 + e_0
    \end{bmatrix}
    \ ,
\end{equation}
with
\begin{equation}\label{abar_ebar_0}
    a_0 = \epsilon \sin \psi_0 \  ,
    \quad
    e_0 = \beta \epsilon (1 - \rho_0) \cos\psi_0  + \beta\epsilon^2 \cos 2\psi_0 \ .
\end{equation}
We note that the perturbation $(\delta\rho, \delta\psi)$ has Gaussian statistics with zero mean. 

In contrast to Ref.~\cite{Martin-et-al}, we explicitly consider the noise in the density dynamics. Using It\^o's Lemma~\cite{Gardiner}, we directly deduce the relevant correlators in Fourier space (other correlators vanish) as
\begin{equation}\label{rhoqrhoq}
\begin{aligned}
    \langle |\delta \rho_q|^2 \rangle &= \frac{ a_0 (D_\rho \beta (q^2- e_0) + D_\psi (\alpha q)^2)}{V\mathcal D} \ ,
    \\
    \langle |\delta \psi_q|^2 \rangle &= \frac{D_\psi \alpha (e_0 - q^2 (1+\alpha) + \beta a_0^2)- D_\rho (\beta a_0)^2}{V\mathcal D} \ ,
\end{aligned}
\end{equation}
where
\begin{equation}
    {\mathcal D} =  \alpha (e_0 - q^2(1+\alpha) ) (q^2 - e_0 - \beta a_0^2 ) \ .
\end{equation}
The correlators [Eq.~\eqref{rhoqrhoq}] are all proportional to the noise amplitudes $(D_\rho, D_\psi)$, as expected. We obtain the expression in real space by integrating over the modes $q$, for instance $\langle\delta\psi^2\rangle = \frac{1}{V} \sum_{q_x,q_y} \langle |\delta \psi_q|^2\rangle \simeq \int \frac{d{\bf q}}{(2\pi)^2} \langle |\delta\psi_q|^2 \rangle$. The fields $(\bar\rho, \bar\psi)$ explicitly appear in the correlators; we restore their dependence on space and time a posteriori when considering their evolution equation, in the same spirit as in Refs.~\cite{Martin-et-al, martin2024}. Therefore, combining Eqs.~\eqref{barrho-0} and~\eqref{rhoqrhoq} yields a closed dynamics for $(\bar\rho, \bar\psi)$.


\subsubsection{Landscape modified by fluctuations}

Our aim is to propose an explicit form for the weak-noise dynamics [Eq.~\eqref{barrho-0}] in terms of $(\bar\rho,\bar\psi)$. Indeed, we want to show that the effect of fluctuations can be cast in terms of a modified landscape, by analogy with the bare dynamics [Eq.~\eqref{rho-modelB-noise}]. To this end, we now assume that the strength of density modulation [Eq.~\eqref{ref_elastic_den}] is small ($\epsilon\ll 1$), and that the noise amplitudes $(D_\rho, D_\psi)$ are of order $\epsilon$.

\begin{figure}
    \includegraphics[width=\columnwidth]{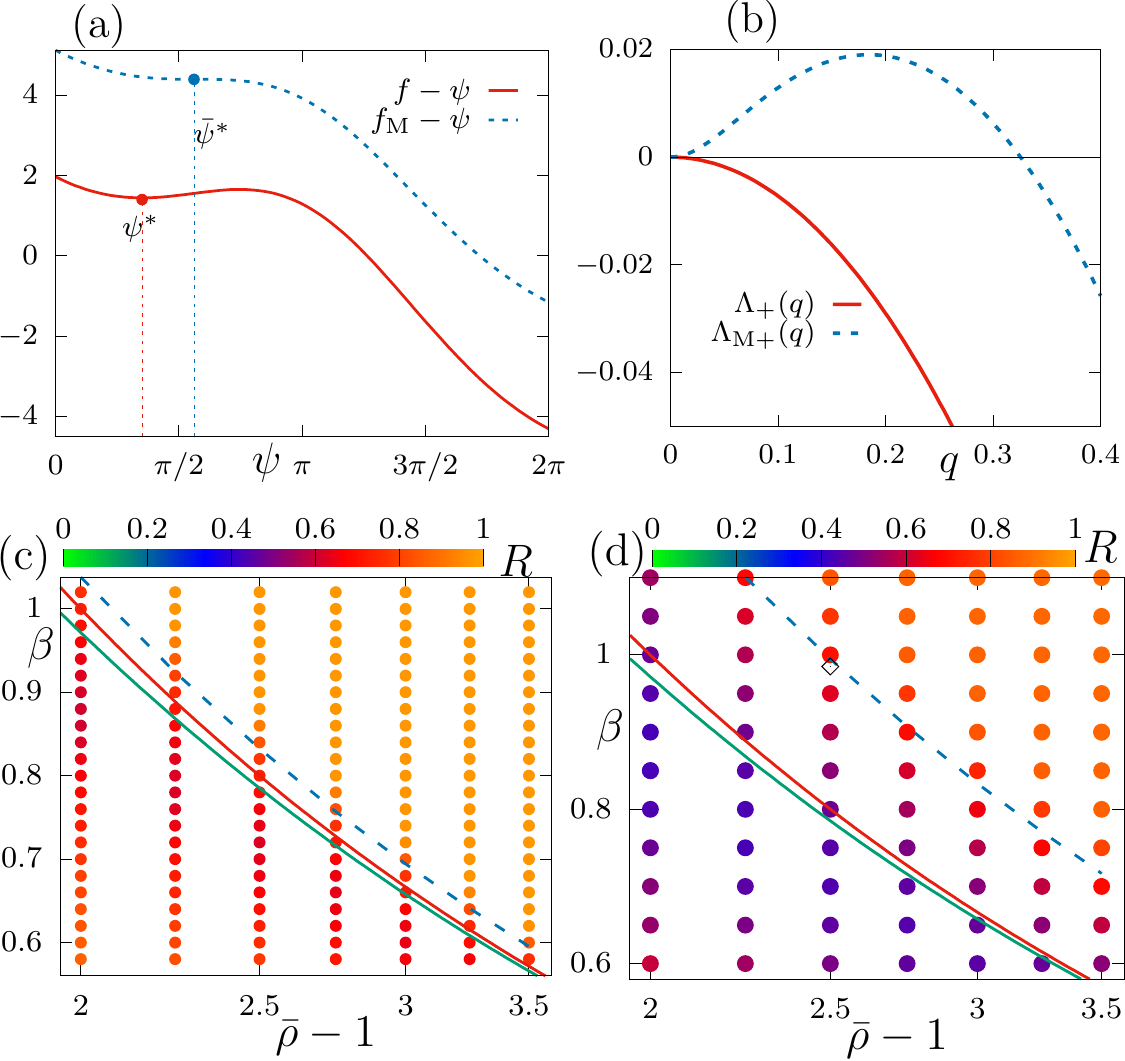}
   \caption{(a)~Comparing the effective landscape $f(\rho,\psi)-\psi$ [Eq.~\eqref{eq:f}] and its counterpart $f_{\rm M}(\rho,\psi)-\psi$ modified at weak noise [Eq.~\eqref{eq:fR_0}] reveals how fluctuations alter the location of the arrested fixed point, respectively at $\bar\psi$ (red) and $\bar\psi_{\rm M}$ (blue). Parameters: $\rho=3.5$, $\beta=0.985$, $\epsilon=0.5$, $D_\psi=D_\rho=0.5$.
    (b)~The eigenvalues for linear stability in the noiseless and weak-noise limits, respectively $\Lambda_{+}$ [Eq.~\eqref{eq:lambda_p}, red] and $\Lambda_{\rm M +}$ [Eq.~\eqref{eq:lambda_p_M}, blue], show that arrest gets destabilized by fluctuations. Same parameters as in (a).
    Phase diagrams in terms of the synchronization parameter $R = \langle | \frac 1 V \int d{\bf r} e^{i \psi} |\rangle$ for $\alpha=1$, $\epsilon=0.5, ~dt=0.001, ~dx=0.25$, (c)~$D_\psi=D_\rho=0.1$ and (d)~$0.5$. The red and green lines  correspond to the stability of the arrested and cycling states, respectively, in the absence of noise, see Fig.~\ref{phase_diagram}. The blue dashed line refers to the stability criterion of the arrested state at weak noise [Eq.~\eqref{noise-corrected-stability line}]. The hollow black diamond corresponds to the parameter values in (a-b).}
\label{analytical-noise-corrected-pd}
\end{figure}

To leading order in $\epsilon$, we deduce that the correlators [Eq.~\eqref{rhoqrhoq}] drastically simplify as
\begin{equation}\label{corrs_2}
    \langle\delta \rho\delta\psi\rangle = O(\epsilon^2) \ ,
    \quad
    \langle\delta\psi^2\rangle = 2 \bar D_\psi + O(\epsilon^2) \ ,
\end{equation}
where
\begin{equation}\label{Dpsibar}
    \bar D _\psi = \frac{D_\psi}{2V} \sum_{q_x,q_y} \frac{1}{q^2} \simeq \frac{D_\psi}{4 \pi} \int_{\frac{2\pi}{L}}^{\frac{\pi}{dx}} \frac{dq}{q},
\end{equation}
where the cut-offs depend on the system size $L$ and the lattice spacing $dx$. The log-divergence of such correlators at large $L$ is a typical feature of two-dimensional systems, which has also been recently reported in other active dynamics at weak noise~\cite{martin2024}. Substituting Eq.~\eqref{corrs_2} into Eq.~\eqref{barrho-0}, it follows that dynamics of $(\rho_0, \psi_0)$ can be written as
\begin{equation}\label{rho-fluc-fin}
    \partial_t \rho_0 = \frac{\alpha}{\beta}\nabla^2 \frac{\partial f_{\rm M}}{\partial\rho_0} \ ,
    \quad
    \partial_t \psi_0 = 1 + \nabla^2 \psi_0 - \frac{\partial f_{\rm M}}{\partial\psi_0} \ ,
\end{equation}
in terms of the modified landscape $f_{\rm M}$ given by
\begin{equation}\label{eq:fR_0}
\begin{aligned}
    f_{\rm M}(\rho_0, \psi_0) &= \beta \epsilon (1-\rho_0) ( 1 - \bar D_\psi) \cos\psi_0
    \\
    &\quad + \frac{\beta \epsilon^2}{4}\cos 2\psi_0 + \frac{\beta}{2} \rho_0^2 + O(\epsilon^3) \ .
\end{aligned}
\end{equation}
Therefore, to leading order, the effect of the noise is to modify the landscape from $f$ [Eq.~\eqref{eq:f}] to $f_{\rm M}$ [Eq.~\eqref{eq:fR_0}]; see Fig.~\ref{analytical-noise-corrected-pd}(a). It remains to explore whether such a modification can potentially destabilize the arrested state.

The arrested state $\bar\psi_{\rm M}$ associated with $f_{\rm M}$ is defined, by analogy with Eq.~\eqref{eq:arr}, as
\begin{equation}
    \frac{\partial f_{\rm M}}{\partial\psi_0}(\bar\psi_{\rm M}) = 1 \ ,
    \quad
    \frac{\partial^2 f_{\rm M}}{\partial\psi_0^2}(\bar\psi_{\rm M}) > 0 \ .
\end{equation}
The stability matrix, given by linearizing the dynamics [Eq.~\eqref{rho-fluc-fin}] around the stationary solution $(\bar\rho, \bar\psi_{\rm M})$, reads
\begin{equation}
    {\mathbb S}_{\rm M}(q) =
    \begin{bmatrix}
        -\alpha q^2 & -\alpha a_{\rm M} \, q^2
        \\
        -\beta a_{\rm M} & - q^2 + e_{\rm M}
    \end{bmatrix}
    \ ,
\end{equation}
and is given in terms of
\begin{equation}
\begin{aligned}
    a_{\rm M} &= \epsilon ( 1- \bar D_\psi) \sin \bar\psi_{\rm M} \ ,
    \\
    e_{\rm M} &= \beta \epsilon (1-\bar\rho) ( 1- \bar D_\psi) \cos \bar\psi_{\rm M}  + \beta \epsilon^2 \cos 2 \bar\psi_{\rm M} \ .
\end{aligned}
\end{equation}
Similarly to the analysis in Sec.~\ref{sec:stab-arr}, it is straight-forward to show that the eigenvalue $\Lambda_{{\rm M +}}(q)$ controlling the destabilization of arrested states has the following form to the lowest order in $q$ :
\begin{equation}\label{eq:lambda_p_M}
    \Lambda_{{\rm M}+}(q) = - \alpha ( 1 + \beta a_{\rm M}^2 / e_{\rm M}) q^2 + O(q^4) \ .
\end{equation}
The full dependence of $\Lambda_{{\rm M}+}(q)$ points at a spinodal-type instability similar to the noiseless case [Eq.~\eqref{eq:lambda_p}]. Yet, the regimes of parameters where such an instability arises are not identical. Interestingly, we can now identify some regimes where the weak-noise analysis predicts the existence of an instability which is missed by its noiseless counterpart [Fig.~\ref{analytical-noise-corrected-pd}(b)]. To systematically quantify such a mismatch, we introduce the modified phase boundary $\beta_{\rm ar, M}(\bar\rho)$, delineating the regime of stability of the arrested state $(\bar\rho, \bar\psi_{\rm M})$, which obeys
\begin{equation}\label{noise-corrected-stability line}
    \beta_{\rm ar, M} a_{\rm M}^2 + e_{\rm M} = 0.
\end{equation}
In practice, comparing $\beta_{\rm ar, M}$ with its noiseless counterpart $\beta_{\rm ar}$ [Eq.~\eqref{pert-pb}] reveals that fluctuations create a wider parameter regime where arrest is unstable; notice the gap between the solid and dotted lines in Figs.~\ref{analytical-noise-corrected-pd}(c,d). To test the validity of our analytical predictions, we obtain the order parameter $R$ [Eq.~\eqref{eq:R}] from numerical simulations of Eq.~\eqref{rho-modelB-noise}. The onset of arrest is given by $R \simeq 1$ [Figs.~\ref{analytical-noise-corrected-pd}(c,d)]. Remarkably, our analytics agrees well with numerical simulations even beyond the regime of weak noise, namely, the transition to $R\simeq 1$ appears close to our predictions (dotted lines) also for larger noise strengths shown in Fig.~\ref{analytical-noise-corrected-pd}(d).


\subsection{Instability of cycling state}\label{sec:floquet}

We now consider the stability of the homogeneous cycling state, with constant density $\bar\rho=\frac 1 V \int_V d{\bf r} \rho$ and time-dependent, periodic phase $\Psi(t)$ that obeys
\begin{equation}\label{dynamical-psi}
    \dot\Psi = 1 - \frac{\partial f}{\partial \psi}(\bar\rho,\Psi) .
\end{equation}
The linearized dynamics of the perturbations around the homogeneous cycling state
\begin{equation}
\begin{aligned}
    \rho_q &= \frac 1 V \int_V d{\bf r} e^{i {\bf q}\cdot{\bf r} } \big[ \rho({\bf r},t) - \bar\rho\big] \ ,
    \\
    \Psi_q &= \frac 1 V \int_V d{\bf r} e^{i {\bf q}\cdot{\bf r} } \big[ \psi({\bf r},t) - \Psi(t)\big] \ ,
\end{aligned}
\end{equation}
is again easily expressed in the Fourier domain as
\begin{equation}
\label{eq:lin_dynamics_cycling}
    \frac{d}{dt}
    \begin{bmatrix}
    \rho_q \\ \Psi_q
    \end{bmatrix}
    = {\mathbb S}_{\rm C}(q,t)
    \begin{bmatrix}
    \rho_q \\ \Psi_q
    \end{bmatrix}
     \ ,
\end{equation}
where the stability matrix
\begin{equation}\label{eq:Sc}
    {\mathbb S}_{\rm C}(q,t) =
    \begin{bmatrix}
    -\alpha q^2 & -\alpha a_{\rm C}(t) q^2
    \\
    - \beta a_{\rm C}(t)  & - q^2 + e_{\rm C}(t)
    \end{bmatrix}
    \ ,
\end{equation}
is given in terms of
\begin{equation}
\begin{aligned}
    a_{\rm C}(t) &= \epsilon \sin \Psi(t) \  ,
    \\
    e_{\rm C}(t) &= \beta \epsilon (1 - \bar\rho) \cos \Psi(t)  + \beta\epsilon^2 \cos 2\Psi(t) \ .
\end{aligned}
\end{equation}
In contrast with the stability matrix ${\mathbb S}_{\rm A}(q)$ of the arrested state [Eq.~\eqref{M0-arrest-dimless}], the stability matrix ${\mathbb S}_{\rm C}(q,t)$ of the cycling state is time-dependent and periodic, with the same period $\tau$ as that of the homogeneous cycling phase $\Psi(t)$ [Eq.~\eqref{dynamical-psi}].

For a dynamical system with periodic coefficients, as in Eq.~\eqref{eq:lin_dynamics_cycling}, the solutions themselves need not be periodic. The general solution can be written as
\begin{equation}\label{mu_i-def}
    ( \rho_q, \Psi_q )  = \sum_{i=1}^2 c_i(q) \, e^{\mu_i(q) t} {\bf p}_i(q,t) \ ,
\end{equation}
where $c_i(q)$ denote constants depending on initial conditions, functions ${\bf p}_i(q,t)$ are vector-valued having period $\tau$, and complex numbers $\mu_i(q)$ are known as {\it Floquet exponents}~\cite{Klausmeier2008}. These exponents represent the growth rates averaged over a limit cycle of  perturbations along the directions given by ${\bf p}_i(q,t)$. In consequence, Floquet exponents $\mu_i(q)$ determine the long-time behavior of $(\rho_q,\Psi_q)$. If the $\mu_i(q)$ all have negative real parts, the solutions are stable; if any $\mu_i(q)$ has a positive real part, the fixed point becomes unstable as then the perturbation will grow indefinitely with time.

Equivalently, one may also analyze the stability of the linearized dynamics [Eq.~\eqref{eq:lin_dynamics_cycling}] by considering the {\em Floquet multipliers} defined as
\begin{equation}
    g_i(q)= e^{\mu_i(q) \tau} .
\end{equation}
The limit cycle is linearly stable if $|{\rm Re}[g_{1,2}(q)]|<1$. To calculate the multipliers $g_i(q)$, we introduce the fundamental matrix ${\mathbb M}(q,t)$~\cite{Klausmeier2008} which obeys 
\begin{equation}
\label{fundamental_matrix_defn}
    \frac{d}{dt} {\mathbb M} (q,t) = {\mathbb S}_{\rm C}(q,t) {\mathbb M}(q,t) \ ,
\end{equation}
with the initial condition ${\mathbb M} (q,0)$ given by the identity matrix. Integrating Eq.~\eqref{fundamental_matrix_defn} over one period, from $t=0$ to $t=\tau$, one obtains the monodromy matrix ${\mathbb M} (q,\tau)$ that describes the evolution of the perturbation $(\rho_q,\Psi_q)$ over one period of the limit cycle. In practice, the multipliers $g_i$ coincide with the eigenvalues of ${\mathbb M} (q,\tau)$~\cite{Kulkarni2020}. In the absence of an analytical solution for $\Psi(t)$, we first numerically integrate Eq.~\eqref{dynamical-psi}, using an Euler discretization scheme, to extract $\tau$ and $\Psi(t)$ for given values of $(\beta, \epsilon)$. We then solve Eq.~\eqref{fundamental_matrix_defn} for a discretized range of values of $q$, which leads to the determination of ${\mathbb M} (q,\tau)$ and its eigenvalues $g_i(q)$ for given parameters $(\beta ,\epsilon,\alpha, \bar\rho)$.

For short-wavelength perturbations, namely sufficiently large values of $q$, the cycling state is stable. Indeed, the stability matrix ${\mathbb S}_{\rm C} (q,t)$ [Eq.~\eqref{eq:Sc}] is dominated by its terms proportional to $-q^2$, so that it reduces to an upper-triangular matrix, and the eigenvalues are simply given by the diagonal terms which are always negative. Any transition to instability can thus only occur for intermediate values of $q$. We report in Fig.~\ref{phase_diagram} that the corresponding prediction for the transition line $\beta_{\rm cy}$ from stable to unstable cycles (black solid line) shows a very good agreement with numerical simulations. Snapshots in Figs.~\ref{pattern-2-fig}(g-l) show the formation and growth of domains, analogous to the spinodal scenario for the instability of arrest [Sec.~\ref{sec:stab-arr}]. When the local density goes below a given threshold, waves start to propagate and undergo sustained pulsation, yielding stationary dynamical patterns [Sec.~\ref{sec:waves}].


\section{Coarse-grained dynamics}\label{app:append-coarse-graining}

In this Section, we demonstrate that the hydrodynamic equations for the density field $\rho$ and the phase field $\psi$ can be obtained in a form similar to Eq.~\eqref{rho-modelB-noise} by coarse-graining a generic model of pulsating particles, inspired by Refs.~\cite{Yiwei-Etienne-PAM, manacorda2023pulsating, Liu_2024, pineros2024biased}.

\subsection{From particles to fields}

We start with a microscopic model where the state of each particle $i$ is described by its position ${\bf r}_i$ and its internal variable $\theta_i$. The latter determines the particle size $\sigma(\theta_i)$ as
\begin{equation}
    \sigma(\theta) = \sigma_0\frac{1+ \lambda \sin \theta}{1+ \lambda} \ ,
\end{equation}
where $\lambda$ and $\sigma_0$ are the oscillation amplitude and the maximum size, respectively. Particles interact through the potential $U({\bf r}_i-{\bf r}_j, \sigma(\theta_i),\sigma(\theta_j))$ that is a function of their separation distance and sizes:
\begin{equation}\label{eq:dyn_mic}
\begin{aligned}
    \dot{{\bf r}}_{i} &= -\mu \partial_{{\bf r}_i} U + \sqrt{2\mu T} \bm{\xi}_i \ ,
    \\
    \dot{\theta}_{i} &= \Omega + \epsilon_0 \sum_{j\in\partial_i} \sin (\theta_j -\theta_i) - \mu_{\theta} \partial_{\theta_i} U + \sqrt{2 \mu_\theta T} \eta_i \ ,
\end{aligned}
\end{equation}
where $\bm{\xi}_i$ and $\eta_i$ represent uncorrelated Gaussian white noises with unit variance and zero mean, $\epsilon_0$ is a synchronization strength, $\mu$ and $\mu_\theta$ are mobilities, $T$ is the temperature, and $\Omega$ is the drive acting on each particle. The sum $\sum_{j\in\partial_i}$ runs over the particles $j$ which are in the vicinity of particle $i$, such that it fixes the finite range of synchronizing interaction. So far, most studies on pulsating particles have considered repulsive interactions~\cite{Togashi2019, Yiwei-Etienne-PAM, manacorda2023pulsating, Liu_2024, pineros2024biased, casagrande2026a, casagrande2026b}. Our coarse-graining accommodates a generic potential $U$ which can have both repulsive and attractive components; for instance, this is the case for interactions as in the vertex model~\cite{farhadifar2007influence, staple2010mechanics, PhysRevX.6.021011}, which describes the dynamics of tissues as an assembly of confluent polygons [Fig.~\ref{model-fig}].

For simplicity, we assume in what follows that the potential $U$ is controlled only by the packing fraction $\phi$, given by
\begin{equation}
    \phi = \frac{\pi}{L^2} \sum_j \sigma^2(\theta_j) \ ,
\end{equation}
where $L$ represents system size. Within this scheme, the quantities $\partial_{{\bf r}_i} U$ and $\partial_{\theta_i} U $ can be approximated as
\begin{equation}\label{eq:int}
\begin{aligned}
    \partial_{{\bf r}_i} U & \simeq (\partial_{\phi} U) \partial_{{\bf r}_i} \sum_j \frac{\pi \sigma^2(\theta_j)}{L^2} \delta ({\bf r}_i - {\bf r}_j) \ ,
    \\
    \partial_{\theta_i} U &\simeq (\partial_{\phi} U) \sum_j \delta ({\bf r}_i - {\bf r}_j) (\partial_{\theta_j}\phi) \ ,
\end{aligned}
\end{equation}
where we assume that $\partial_{\phi} U$ is a constant, independent of particle coordinates $({\bf r}_i,\theta_i)$, at mean-field level. These assumptions amount to mapping the effect of pairwise interactions into some one-body, external fields applied on positions and phases. Then, the dynamics in Eq.~\eqref{eq:dyn_mic} becomes
\begin{equation}\label{pam-rdot}
\begin{aligned}
    \dot{{\bf r}}_{i} &= - k \, \partial_{{\bf r}_i} \sum_j (1+ \lambda \sin \theta_j)^2 \, \delta ({\bf r}_i - {\bf r}_j) + \sqrt{2\mu T} \bm{\xi}_i \ ,
    \\
    \dot{\theta}_{i} &= \Omega + \sum_j \left[ \epsilon_0 \sin(\theta_j - \theta_i) - c \cos \theta_j - \frac{\lambda c}{2} \right] \delta({\bf r}_i -{\bf r}_j)
    \\
    &\quad + \sqrt{2 \mu_\theta T} \eta_i \ ,
\end{aligned}
\end{equation}
where
\begin{equation}
    k = \frac{\mu \pi \sigma_0^2 (\partial_\phi U)}{L^2(1+\lambda)^2} \ ,
    \quad
    c = \frac{2 \pi \sigma_0^2 \lambda \mu_\theta (\partial_\theta U)}{L^2(1+\lambda)^2} \ .
\end{equation}    
Note that we have replaced $\sum_{j\in\partial_i}$ with $\sum_j \delta({\bf r}_i-{\bf r}_j)$, and also used $\delta({\bf r}_i-{\bf r}_j)$ in Eq.~\eqref{eq:int}. This choice is justified if one assumes that the range of interactions is small compared to the spatial variations of the fields at the hydrodynamic level~\cite{PhysRevLett.108.248101}.

To derive a coarse-grained description of the system dynamics, we first need to obtain the dynamics of the empirical probability density function given by
\begin{equation}
\begin{aligned}
    P({\bf r}, \theta,t) &= \sum_i P_i({\bf r}, \theta,t) \ ,
    \\
    P_i({\bf r}, \theta,t) &= \delta({\bf r} - {\bf r}_i(t)) \delta(\theta - \theta_i(t)) \ .
\end{aligned}
\end{equation}
Note that $P$ is a stochastic density for a single realization. We follow the Dean-Kawasaki procedure~\cite{dean, kawasaki} to handle the set of stochastic equations of motion [Eq.~\eqref{pam-rdot}]. Using It\^o's lemma~\cite{Gardiner}, we have
\begin{equation}
    \partial_t P = \sum_i (\dot{\bf r}_i \cdot \partial_{{\bf r}_i} + \dot{\theta}_i \partial_{\theta_i} + \mu T \partial_{{\bf r}_i{\bf r}_i}^2 + \mu_\theta T \partial_{\theta_i \theta_i}^2 ) P_i \ ,
\end{equation}
from which we deduce
\begin{widetext}
\begin{equation}
\begin{aligned}
    \partial_t P &=T (\mu \nabla^2 + \mu_\theta \partial^2_{\theta \theta}) P - \sum_i (\sqrt{2\mu T} \bm{\xi}_i \cdot \partial_{\bf r} + \sqrt{2 \mu_\theta T} \eta_i \partial_{\theta})P_i  + k \nabla \cdot \bigg[ P({\bf r},\theta,t) \nabla \sum_j (1+ \lambda \sin \theta_j)^2 \delta({\bf r}-{\bf r}_j) \bigg]
    \\
    &\quad - \partial_{\theta}\bigg\{ P({\bf r},\theta, t)\bigg[\Omega + \sum_j ( \epsilon_0 \sin(\theta_j -\theta) - c\cos \theta_j - (\lambda c/2) \sin2 \theta_j ) \delta({\bf r}- {\bf r}_j) \bigg]\bigg\} \ ,
\end{aligned}
\end{equation}
yielding
\begin{equation}\label{r-term-fin}
\begin{aligned}
    \partial_t P &= T(\mu \nabla^2 + \mu_\theta \partial^2_{\theta \theta})P - \sum_i ( \sqrt{2\mu T} \bm{\xi}_i \cdot \partial_{\bf r} + \sqrt{2 \mu_\theta T} \eta_i \partial_{\theta}) P_i
    \\
    &\quad - \partial_{\theta}\bigg\{ P({\bf r},\theta, t) \bigg[ \Omega + \int d\theta' ( \epsilon_0 \sin(\theta' -\theta) - c\cos \theta' - (\lambda c/2) \sin 2 \theta' ) P({\bf r},\theta', t) \bigg] \bigg\}
    \\
    &\quad + k \nabla \cdot \bigg[ P({\bf r},\theta, t)  \nabla \int d \theta' (1+ 2 \lambda \sin \theta' + \lambda^2 \sin^2 \theta') P({\bf r}, \theta',t) \bigg] \ .
\end{aligned}
\end{equation}
The last term is the only new contribution to the evolution equation of the empirical probability density function when compared with the result obtained in Ref.~\cite{Yiwei-Etienne-PAM}, which neglected the contribution of $U$ in the dynamics of ${\bf r}$ at the coarse-grained level.

Next, we introduce the harmonics 
\begin{equation}
    f_n({\bf r},t) = \int d\theta e^{i n \theta} P({\bf r}, \theta, t) \ .
\end{equation}
From Eq.~\eqref{r-term-fin}, we deduce the dynamics of $f_n$ as
\begin{equation}\label{fn-eom}
\begin{aligned}
    \partial_t f_n &= in \Omega f_n + T(\mu \nabla^2 - \mu_\theta n^2 ) f_n + \frac{n \epsilon_0}{2}(f_{n-1}f_1 - f_{n+1}f_{-1}) - i n c f_n {\rm Re}[f_1] +\frac{i n \lambda c }{2} f_n {\rm Im}[f_2]
    \\
    &\quad + k \nabla \cdot \big[ f_n \nabla ( (1+\lambda^2/2) f_0 + 2 \lambda {\rm Im}[f_1] -(\lambda^2/2){\rm Re}[f_2]) \big] \ ,
\end{aligned}
\end{equation}
where we have neglected the noise terms for simplicity. Thus, the dynamics of each mode $f_n$ depends on the higher order ones $f_{n+1}$. For the first three modes, we get
\begin{equation}\label{f0-dyn}
\begin{aligned}
    \partial_t f_0 &= \mu T \nabla^2 f_0 + k \nabla \cdot\big[ f_0 \nabla ( (1+\lambda^2/2) f_0 + 2 \lambda {\rm Im}[f_1] - (\lambda^2/2){\rm Re}[f_2] ) \big] \ ,
    \\
    \partial_t f_1 &= i \Omega f_1 +  T(\mu \nabla^2 - \mu_\theta) f_1 + \frac{\epsilon_0}{2}(f_0f_1-f_2 f_{-1}) - i c ({\rm Re}[f_1] + (\lambda/2){\rm Im}[f_2] ) f_1
    \\
    &\quad + k \nabla \cdot \big[ f_1 \nabla ( (1+\lambda^2/2) f_0 + 2 \lambda {\rm Im}[f_1] -(\lambda^2/2) {\rm Re}[f_2] ) \big] \ ,
    \\
    \partial_t f_2 &= 2i \Omega f_2 +  T(\mu \nabla^2 - 4 \mu_\theta) f_2 + \epsilon_0 (f_1^2-f_3f_{-1}) - 2i c ({\rm Re}[f_1] + (\lambda/2){\rm Im}[f_2] ) f_2 
    \\
    &\quad +  k \nabla \cdot \big[ f_2 \nabla ( (1+\lambda^2/2) f_0 + 2 \lambda {\rm Im}[f_1] -(\lambda^2/2) {\rm Re}[f_2] ) \big] \ .
\end{aligned}
\end{equation}
We now assume a scaling of the operators and harmonics, with respect to a small number $\chi$, as $(\partial_t, \nabla^2) \sim \chi^2$ and $f_n \sim |\chi|^n$. Considering that $f_2$ relaxes fast allows us to express $f_2$ in terms of $f_1$ to leading order~\cite{Yiwei-Etienne-PAM} as
\begin{equation}\label{f2-f1}
    f_2 \simeq \frac{\epsilon_0 f_1^2}{4 \mu_\theta T - 2 i \Omega} \ .
\end{equation}
The dynamics of the density field $\rho=f_0$ and the complex field $A=f_1$ can be obtained upon substituting Eq.~\eqref{f2-f1} into Eq.~\eqref{f0-dyn}. For the density field, we obtain
\begin{equation}\label{rho-coarse-hydro}
    \partial_t \rho = \mu T \nabla^2 \rho +  k_1 \nabla \cdot (\rho \nabla \rho) - i k_2 \nabla \cdot [\rho \nabla(A - \bar A)] \ ,
\end{equation}
where $k_1= k(1+\lambda^2/2)$, $k_2 = \lambda k$, and $\bar A$ denotes the complex conjugate of $A$. In the above equation we have ignored the higher order sub-dominant contribution from the $f_2$ term in the dynamics of $f_0$. For the complex field, we obtain
\begin{equation}\label{A-coarse-hydro}
\begin{aligned}
    \partial_t A &= (i \Omega + T \mu \nabla^2 - \mu_\theta T )A + \frac{\epsilon_0}{2} \rho A - \frac{\epsilon_0^2(2\mu_\theta T +  i \Omega) |A|^2}{16T^2\mu_\theta^2 + 4\Omega^2}A - \frac{ic}{2}A(A+\bar A)
    \\
    &\quad - \frac{i c \lambda \epsilon_0 A}{16T^2\mu_\theta^2 + 4\Omega^2} {\rm Im}[(2\mu_\theta T +  i \Omega)A^2] + k_1 \nabla\cdot(A\nabla \rho) \ .
\end{aligned}
\end{equation}
Equations~\eqref{rho-coarse-hydro} and~\eqref{A-coarse-hydro} give the coarse-grained hydrodynamic description for a collection of interacting, pulsating particles at the noise-free level. Decomposing $A=R e^{i \psi}$ in terms of the hydrodynamic amplitude $R$ and phase $\psi$, we obtain
\begin{equation}\label{R-coarse}
\begin{aligned}
    \partial_t R &= \mu T \nabla^2 R - \mu_\theta T R +\frac{\epsilon_0}{2} \rho R + k_1 R \nabla^2 \rho + k_1 (\nabla R)\cdot(\nabla \rho) - \mu T R (\nabla \psi)^2 - \alpha_1 R^3 \ ,
    \\
    \partial_t \psi &= {\Omega} +\frac{2\mu T}{R}(\nabla R)\cdot(\nabla \psi)  + \mu T \nabla^2 \psi - c \cos \psi + k_1 (\nabla \rho)\cdot(\nabla \psi) - R^2 (\alpha_2 + \beta_1 \cos 2\psi + \beta_2 \sin 2\psi) \ ,
\end{aligned}
\end{equation}
where 
\begin{equation}
    \alpha_1 = \frac{\epsilon_0^2\mu_\theta T}{8T^2\mu_\theta^2 + 2\Omega^2} \ ,
    \quad
    \alpha_2 = \frac{\epsilon_0^2 \Omega}{16T^2\mu_\theta^2 + 4\Omega^2} \ ,
    \quad
    \beta_1 = \frac{c \lambda \epsilon_0 \Omega}{16T^2\mu_\theta^2 + 4\Omega^2} \ ,
    \quad
    \beta_2 = \frac{2 \mu_\theta T}{16T^2\mu_\theta^2 + 4\Omega^2} .
\end{equation}
Equation~\eqref{R-coarse} gives the dynamics of the amplitude and phase of the complex field $A$. Coupled with Eq.~\eqref{rho-coarse-hydro}, it provides the coarse-grained dynamics of the pulsating system.


\subsection{Comparison with hydrodynamic model}

We now discuss the conditions under which the coarse-grained dynamics in Eqs.~\eqref{rho-coarse-hydro} and~\eqref{R-coarse} compares with the (noiseless) phenomenological dynamics for the density and phase fields [Eq.~\eqref{rho-modelB-noise}]. Describing the pulsating system through Eq.~\eqref{rho-modelB-noise} assumes that the amplitude $R$ reduces to a constant without any variation in space and time. In practice, this assumption amounts to enforcing that the particles are synchronized uniformly throughout the system. Enforcing the condition of vanishing gradients in Eq.~\eqref{R-coarse} yields
\begin{equation}\label{R-approx}
	R \simeq \frac{\sqrt{\frac{\epsilon_0}{2} \rho - \mu_\theta T} }{ \sqrt{\alpha_1 } } ,
\end{equation}
with the condition $\rho > 2\mu_\theta T/\epsilon_0$. Substituting Eq.~\eqref{R-approx} into Eqs.~\eqref{R-coarse} and~\eqref{rho-coarse-hydro} leads to a closed dynamics for $\rho$ and $\psi$:
\begin{equation}\label{rho-reduced}
\begin{aligned}
    \partial_t \rho &= \mu T \nabla^2 \rho +  k_1 \nabla \cdot (\rho \nabla \rho) + 2 \frac{k_2}{\sqrt{\alpha_1}} \nabla \cdot \left[\rho \nabla \left(\sin \psi \sqrt{\frac{\epsilon_0 \rho}{2}- \mu_\theta T}\right)\right] \ ,
    \\
    \partial_t \psi &= \Omega + \frac{\alpha_2 \mu_\theta T}{\alpha_1} -\frac{\alpha_2 \epsilon_0}{2 \alpha_1}\rho + \mu T \nabla^2 \psi  - c \cos \psi + k_1 (\nabla \rho)\cdot(\nabla \psi) - \frac{1}{\alpha_1} \left(\frac{\epsilon_0 \rho}{2}-\mu_\theta T \right) ( \beta_1\cos 2\psi  + \beta_2\sin 2\psi) \ .
\end{aligned}
\end{equation}
\end{widetext}
We now argue that the coarse-grained [Eq.~\eqref{rho-modelB-noise}] and phenomenological [Eq.~\eqref{rho-reduced}] dynamics belong to the same class of hydrodynamic models. First, both dynamics break the invariance under phase shift, which is a signature of pulsating active matter~\cite{Yiwei-Etienne-PAM, manacorda2023pulsating, pineros2024biased}: a homogeneous phase shift $\psi\to\psi+C$, with arbitrary constant $C$, changes the time-evolution of $\psi$. Therefore, for a sufficiently low drive $\Omega$, Eq.~\eqref{rho-reduced} admits a fixed point $(\bar\rho, \bar\psi_{\rm CG})$, where $\bar\rho=\frac 1 V\int_V d{\bf r}\rho$ is the total density, which corresponds to an arrested state analogous to that of Eq.~\eqref{rho-modelB-noise}. Second, the evolution of the perturbations in the Fourier domain
\begin{equation}
\begin{aligned}
    \rho_q &= \frac 1 V \int_V d{\bf r} e^{i {\bf q}\cdot{\bf r} } \big[ \rho({\bf r},t) - \bar\rho\big] \ ,
    \\
    \psi_q &= \frac 1 V \int_V d{\bf r} e^{i {\bf q}\cdot{\bf r} } \big[ \psi({\bf r},t) - \bar\psi_{\rm CG}\big] \ ,
\end{aligned}
\end{equation}
takes the following form to linear order:
\begin{equation}
\frac{d}{dt}
\begin{bmatrix}
    \rho_q \\ \psi_q
\end{bmatrix}
= {\mathbb S}_{\rm CG}(q) 
\begin{bmatrix}
    \rho_q \\ \psi_q
\end{bmatrix}
\end{equation}
where the stability matrix
\begin{equation}
    {\mathbb S}_{\rm CG}(q) =
    \begin{bmatrix}
        -{\cal A} q^2 & - {\cal B} q^2
        \\
        -{\cal C} &  - \mu T q^2 + {\cal E}
    \end{bmatrix}
    \ ,
\end{equation}
is given in terms of
\begin{equation}
\begin{aligned}
\mathcal{A} &= \mu T  \, + \, k_1 \bar\rho + \, \frac{k_2\sin \bar\psi_{\rm CG} \sqrt{\epsilon_0 \bar\rho}}{\sqrt{2 \alpha_1}} \left(1+ \frac{\mu_\theta T}{\epsilon_0 \bar\rho} \right) \ ,
\\
\mathcal{B} &= k_2 \cos \bar\psi_{\rm CG} \left( \bar\rho - \frac{\mu_\theta T}{\epsilon_0} \right) \sqrt{\frac{2 \epsilon_0 \bar\rho}{\alpha_1}} \ ,
\\
\mathcal{C} &= \frac{\epsilon_0 (\alpha_2 + \beta_1 \cos 2\bar\psi_{\rm CG} + \beta_2 \sin 2 \bar\psi_{\rm CG})}{2 \alpha_1} \ ,
\\
\mathcal{E} &= \frac{2}{\alpha_1}\left( \frac{\epsilon_0 \bar\rho}{2}- \mu_\theta T \right)(\beta_1 \sin 2\bar\psi_{\rm CG} - \beta_2 \cos 2 \bar\psi_{\rm CG}) \ .
\end{aligned}
\end{equation}
The stability matrix ${\mathbb S}_{\rm CG}$ controlling the linearized version of the phenomenological dynamics [Eq.~\eqref{rho-reduced}] has the same qualitative $q-$dependence as its counterpart [Eq.~\eqref{M0-arrest-dimless}] for the phenomenological dynamics [Eq.~\eqref{rho-modelB-noise}]. This shows that the coarse-grained and phenomenological models entail similar linear instabilities. Note that one can also actually obtain a nonlinear contribution of the form $(\nabla \rho)\cdot(\nabla \psi)$ in the phenomenological dynamics of $\psi$ [Eq.~\eqref{rho-modelB-noise}] by admitting a density-dependent interface parameter $\kappa(\rho) = \kappa_0 + \kappa_1 \rho$ in the free energy [Eq.~\eqref{comb-Free-en}]. Similar field-dependent phenomenological parameters have previously been considered in studies on asymmetric~\cite{david-membrane1} and symmetric~\cite{tirtha-membrane1, tirtha-membrane2} fluid membranes.


\section{Discussion}\label{sec:summ}

Using a phenomenological approach, we propose a hydrodynamic theory for a confluent system of pulsating particles [Fig.~\ref{model-fig}]. Our theory relies on some minimal, generic assumptions which encompass a broad class of systems. We argue that the mechanical coupling between the local density and the local phase of particles, which results from the conservation of particle number and of momentum, can be described in terms of a free energy. We capture the existence of an internal pulsation by considering that an autonomous drive is applied to the local phase. Such a top-down perspective is reminiscent of how different types of field theories have been built to capture the physics of many soft and living systems~\cite{hohenberg-review, Bray1994, active-matter-marchetti}. Importantly, we show that our phenomenological approach is actually consistent with the coarse-graining of microscopic models of pulsating particles [Sec.~\ref{app:append-coarse-graining}].

The competition between the drive and the free energy yields three distinct states: homogeneous cycles, homogeneous arrest, and pulsating patterns. Arrest arises due to breakdown of invariance under phase shift~\cite{Yiwei-Etienne-PAM, manacorda2023pulsating, pineros2024biased, casagrande2026a, casagrande2026b}, while cycles and pulsating patterns illustrate the breakdown of time-translational invariance~\cite{sriram-prl-2013}. Owing to the simplicity of our model, we obtain analytical predictions for the phase boundaries in parameter space [Fig.~\ref{phase_diagram}], and examine how fluctuations affect these boundaries while retaining the same phenomenology [Fig.~\ref{analytical-noise-corrected-pd}]. The linear analysis, confirmed by numerical simulations, predicts that homogeneous states are destabilized through a spinodal-type scenario [Fig.~\ref{Lam_pm}]. It yields the formation of various domains growing in a background which is either cycling or arrested [Fig.~\ref{pattern-2-fig}]. When the domain growth exceeds a threshold, waves emerge and propagate in the system, through a mechanism reminiscent of secondary instabilities~\cite{Sakaguchi-PTP93, Coullet-Iooss-PRL90}. These waves lead to sustained pulsating patterns in the profiles of density and phase. Instead, in the passive limit ($\omega=0$), the system relaxes to a homogeneous configuration without any current.

Our theory is clearly distinct from other studies, which also attempt to capture dynamical patterns in dense tissues~\cite{Serra-Picamal-Natphys2012,Zaritsky-PLoS, Banerjee-Marchetti-PRL2015,tlili-RSoc18,Peyret-oscillations-waves19,Petrolli-PRL2019,Hino-DevCell2020,Armon-PNAS,Myers2017,Young1997,Xu2015,Oates2012}, in a few important ways. First, we discard the role of self-advection, in contrast with many hydrodynamic studies of active particles~\cite{active-matter-marchetti}, which amounts to ignoring the individual self-propulsion. This assumption is justified in tissues without any net flow of materials~\cite{karma-annrev, Elad2017}, namely when cells are so jammed that their migration is negligible, as in some epithelia~\cite{Hannezo-glassy-tissue}. Second, some studies consider that pulsatile tissues operate out-of-equilibrium due to an active stress that does not derive from a free-energy~\cite{Vijay-PRL, Banerjee2017, Staddon-PlosOne2022}. Instead, we assume that all mechanical forces derive from a free energy, so that nonequilibrium properties here stem from the autonomous drive applied to the local phase. In that respect, our approach combines mechanical principles, given by conservation laws, with phenomenological arguments, inspired by synchronization theory~\cite{Ritort2005, aranson-review}.

Some field theories may, at first glance, appear similar to ours: they feature patterns for a density (conserved) field and a phase (non-conserved) field, despite the lack of self-advection and active stress. We now argue that our theory actually stands out as a unique perspective on pulsating liquids for several reasons. First, some works consider the effect of coupling the CGLE, which describes the evolution of a complex field near a limit cycle~\cite{aranson-review}, with a conserved scalar field~\cite{CoulletandFauve1985, Coullet-Iooss-PRL90, Sakaguchi-PTP93, MatthewsandCox-Nonlinearity2000}. A related body of work studies the hydrodynamics of diffusive oscillators~\cite{tirtha-oscillator,astik-oscillator, Astik-PRELetter}, and also recently in the context of confluent tissues~\cite{li2024fluidization}, in terms of the coupling of density and phase fields. All these theories entail the invariance under phase shift, whereas our model breaks this invariance: therefore, these previous works do not capture the existence of an arrested state, which is an essential feature of pulsating liquids~\cite{Yiwei-Etienne-PAM, manacorda2023pulsating, pineros2024biased}. Second, some studies address the formation of reaction-diffusion patterns in deformable media~\cite{Keldermann2007, Panfilov2005}, yet they do not report the emergence of any arrest. Third, other works examine the coupling between density and phase fields {\em without} invariance under phase shift~\cite{PhysRevE.87.024901, Sakaguchi-Maeyama}. Such theories lead to propagating waves in the phase profile, while the density profile exhibits aggregation at specific locations. In contrast, our theory features waves in the profiles of both phase and density, as observed in pulsating tissues~\cite{karma-annrev, Elad2017, Brunet-review}. In particular, our results could be useful in studies of embryonic cardiac tissues where mechanical signaling dominates over electrical signaling for coordinating heartbeats~\cite{LiuPNAS, Gaetani}. Other interesting experimental platforms relevant to the work presented here could be those involving inflatable particles~\cite{pashine2025}.

We now argue why topological defects cannot {\em spontaneously} emerge in the profiles of density and phase. Our model can be regarded as describing the collective dynamics of particles pulsating close to an ordered state, where particles are perfectly synchronized. In this regime, the phase approximation typically holds~\cite{Sakaguchi-Maeyama}; in other words, we implicitly assume that the amplitude of the complex field [Sec.~\ref{app:append-coarse-graining}], measuring the local degree of synchronization~\cite{aranson-review}, is held fixed. A consequence of this assumption is that our model cannot capture the spontaneous emergence of 
topological defects, which appear by definition at locations where the amplitude of the complex field vanishes. In a similar fashion, the Leslie-Ericksen theory~\cite{Ericksen1991, Leslie1968} of liquid crystals, which assumes that the nematic field has a fixed norm, does not entail any spontaneous defect formation, whereas the Landau-de Gennes theory~\cite{prost1993, EMMRICH201832} contains defects by allowing variations of the amplitude of the nematic field. In our pulsating liquids, relaxing the assumption of a constant amplitude for the complex field could lead to patterns with spiral waves and/or defect turbulence, as reported for pulsating particles~\cite{Yiwei-Etienne-PAM, manacorda2023pulsating}. It would be interesting to explore how the statistics of such defects at the hydrodynamic level compares with its particle-based counterparts. Other interesting open avenues include investigating the effects of inhomogeneous driving~\cite{PhysRevLett.134.188301, 10.3389/fphy.2022.976515} or the inclusion of residual stresses~\cite{PhysRevLett.129.048102} in our model.

\acknowledgments{We thank Alessandro Manacorda for insightful discussions, and Michael E. Cates for comments on the manuscript. \'E.F., T.D. and T.B. were supported through the Luxembourg National Research Fund (FNR), grant references 14389168 and C22/MS/17186249.}


\appendix

\section{Viscous effects}\label{viscosity}

Most tissues have a visco-elastic mechanics~\cite{tlili2015colloquium}. In framing our model equations, we consider the elastic stress to comprise only the isotropic bulk stress and neglect all viscous contributions. Instead, one can model the stress to have both isotropic and anisotropic parts. In practice, it consists of replacing $\Sigma$ in Eq.~\eqref{ext-force-balance} by the total stress
\begin{equation}\label{stress-dissociation}
 \Sigma^{\rm tot}_{ij} = \Sigma'_{ij} + \tilde \Sigma_{ij} \ ,
\end{equation}
with
\begin{equation}\label{iso-stress}
\begin{aligned}
    \Sigma'_{ij} &= \Sigma_{ij} + \overline{\eta} \, \nabla_k v_k \, \delta_{ij} \ ,
    \\
   \tilde \Sigma_{ij} &=  \eta \left[ \left( \nabla_i v_j + (\nabla_i v_j)^{\rm T} \right) - (\nabla_k v_k) \delta_{ij} \right] \ ,
\end{aligned}
\end{equation}
where $\Sigma_{ij}$ follows Eq.~\eqref{eq:stress}. Note that Eq.~\eqref{stress-dissociation} expresses the total stress as a combination of a diagonal and a traceless matrix. Also, $\overline{\eta}$ and $\eta$ correspond to bulk and shear viscosity, respectively. Accordingly, at zero noise, combining Eqs.~\eqref{ext-force-balance} and~\eqref{iso-stress} gives the required modified condition for stress balance:
\begin{equation}\label{v-balance}
\begin{aligned}
     v_i &= -\frac{\lambda}{\gamma \rho_0 \rho} \nabla_i \rho - \frac{\epsilon \lambda}{\gamma \rho} \sin \psi \nabla_i \psi + \frac{\overline{\eta}}{\gamma \rho} \, \nabla_i \nabla_k  v_k
    \\
    &\quad + \frac{\eta}{\gamma \rho} \nabla_j \left[(\nabla_i v_j + (\nabla_i v_j)^{\rm T}) -  (\nabla_k v_k) \delta_{ij}\right] \ .
\end{aligned}
\end{equation}
The presence of anisotropic terms in \eqref{v-balance} makes the analysis considerably more challenging than the one presented in the main text.


\section{Hydrodynamics of density and magnetization}
\label{sec:current_closed}

In this Appendix, we derive the closed dynamics for the density field $\rho$ and the magnetization field $\bf m$:
\begin{equation}
    {\bf m} = \epsilon a \nabla ( \cos \psi) \ ,
\end{equation}
defined so that
\begin{equation}
    \partial_t \rho = a \nabla^2 \rho - \nabla \cdot {\bf m} \ ,
\end{equation}
where $a = \lambda/(\gamma \rho_0)$, and we take non-dimensionalized units of density by scaling $\rho$ with $\rho_0$. The evolution equation for ${\bf m}$ follows as
\begin{equation}\label{m_defn5}
\begin{aligned}
    \partial_t {\bf m} &= -\epsilon a  \nabla \big[  \omega \sin \psi + D \sin \psi \nabla^2 \psi
    \\
    &\quad + b \sin^2 \psi (1+ \epsilon \cos \psi - \rho) \big] \ ,
\end{aligned}
\end{equation}
where $b =\mu \lambda \epsilon$, from which we deduce
\begin{equation}\label{m_i-t_terms}
    \partial_t m_i = \epsilon a \big[ {\mathcal T}_1 \nabla_i \psi + {\mathcal T}_2 (\nabla^2 \psi) \nabla_i \psi + {\mathcal T}_3\nabla_i\nabla^2 \psi + {\mathcal T}_4 \nabla_i \rho \big] ,
\end{equation}
where 
\begin{equation}\label{T4_defn1}
\begin{aligned}
    {\mathcal T}_1 &= - \omega \cos \psi + b \epsilon \sin^3 \psi - b \sin 2 \psi (1+ \epsilon \cos \psi-\rho) \ ,
    \\ 
    {\mathcal T}_2 &= -D \cos \psi \ ,
    \quad
    {\mathcal T}_3 = - D \sin \psi \ ,
    \quad
    {\mathcal T}_4 = b \sin^2 \psi \ . 
\end{aligned}
\end{equation}
We express the gradients of $\psi$ as
\begin{equation}
    \nabla_i \psi = \frac{-m_i}{\epsilon a \sin \psi} \ ,
    \quad
    \nabla^2 \psi = -\frac{1}{\epsilon a} \left[ \frac{\nabla \cdot {\bf m}}{\sin \psi} + \frac{|{\bf m}|^2 \cos \psi}{\epsilon a \sin^3 \psi}  \right] \ ,
\end{equation}
and
\vspace{0.2cm}

\begin{widetext}
\begin{equation}\label{t3-term}
\begin{aligned}
   {\mathcal T}_2 (\nabla^2 \psi) \nabla_i \psi &= -\frac{D}{(\epsilon a)^2} \left[\frac{(\nabla \cdot {\bf m}) \cos \psi}{\sin^2 \psi} + \frac{|m|^2 \cos^2 \psi}{\epsilon a \sin^4 \psi} \right]m_i \ ,
   \\
   {\mathcal T}_3 \nabla_i (\nabla^2 \psi) &= \frac{D \sin \psi}{\epsilon a} \left[\frac{\nabla_i (\nabla \cdot {\bf m})}{\sin \psi} + \frac{(\nabla \cdot {\bf m}) m_i \cos \psi }{\epsilon a \sin^3 \psi} + \frac{2 m_j (\nabla_i m_j) \cos \psi}{\epsilon a \sin^3 \psi} + m^2 m_i\left(\frac{1}{\epsilon^2 a^2 \sin^3 \psi} + \frac{3 \cos^2 \psi}{\epsilon^2 a^2 \sin^5 \psi} \right) \right]  \ ,
\end{aligned}
\end{equation}
\end{widetext}
yielding
\begin{equation}\label{mi_evolution_exact}
\begin{aligned}
    \partial_t m_i &= D \nabla_i (\nabla \cdot {\bf m})  + {\mathcal D}_0 m_j \nabla_i m_j + {\mathcal C}_0 \nabla_i \rho
    \\
    &\quad + {\mathcal A}_0 m_i+ {\mathcal E}_0 \rho m_i + {\mathcal F}_0 m^2 m_i \ ,
\end{aligned}
\end{equation}
where
\begin{equation}\label{F0}
\begin{aligned}
    {\mathcal A}_0 &=  \omega \cot \psi - b \epsilon  \sin^2\psi + 2 b \cos \psi(1+ \epsilon \cos \psi) \ ,
    \\
    {\mathcal D}_0 &= \frac{2 D \cos \psi}{\epsilon a \sin^2 \psi} \ ,
    \quad
    {\mathcal F}_0 = \frac{D}{\epsilon^2 a^2} \frac{1+ \cos^2 \psi}{\sin^4 \psi} \ ,
    \\
    {\mathcal E}_0 &= -2 b \cos \psi \ ,
    \quad
    {\mathcal C}_0 = \epsilon a b \sin^2 \psi \ .
\end{aligned}
\end{equation}
To close the dynamics, we use the relations
\begin{equation}\label{eq:m-psi}
\begin{aligned}
    \cos \psi &= \frac{1}{\epsilon a} \nabla^{-2} (\nabla \cdot {\bf m}) \ ,
    \\
    \sin \psi &= \sqrt{1 - \cos^2\psi} \quad {\rm if} \; {\rm mod}(\psi, 2\pi) = [0, \pi] \ ,
    \\
    \sin \psi &= - \sqrt{1 - \cos^2\psi} \quad {\rm if} \; {\rm mod}(\psi, 2\pi) = [\pi,2\pi] \ .
\end{aligned}
\end{equation}
In contrast with models of flocking~\cite{Chate2020}, where the dynamics is stabilized by $m^3$ terms, such a term in Eq.~\eqref{mi_evolution_exact} is actually destablizing since ${\cal F}_0>0$. Instead, we expect that the stabilization here stems from the non-local relation between $\psi$ and $\bf m$ [Eq.~\eqref{eq:m-psi}]. 



\section{Numerical methods}
\label{app:num}

We consider a two-dimensional square lattice with $100 \times 100$ sites. In all our simulations, we use grid spacings $dx=dy=0.25$, and the time increment $dt=0.001$ is chosen to be sufficiently smaller than $dx^2/\alpha$. We employ Euler integration and update the lattice sites in parallel. We use $D_\rho=(\alpha/\beta)D_\psi$, which amounts to enforcing that the noise strengths obey the same proportionality relation as they would in thermal equilibrium [Sec.~\ref{sec:fields}]. We choose initial conditions with homogeneous density $\rho({\bf r},t=0)=\bar\rho$, whereas the phase $\psi({\bf r},t=0)$ is the sum of a homogeneous profile, sampled by the uniform distribution in $(0,2\pi)$, and a small perturbation sampled independently at each site by the uniform distribution in $(0,0.1)$. To measure the averages of some observables, we consider $128$ independent realizations for each point in the phase diagram. The observation time $t_o$ is chosen to be ${\cal O}(100 t_c)$ [Eq.~\eqref{eq:scale}].

The discretized version of the Laplacian and divergence operators must respect the isotropy reflected in our model [Eq.~\eqref{rho-modelB-noise}]. To ensure this condition, we take the discretization weights of the $D2Q9$ model on a square lattice~\cite{thampi-ronojoy}. The expression of the discrete Laplacian, when applied to any scalar function $H_{ij}$ (indices $i,j$ refer to the $2d$ lattice sites), then reads
\begin{equation}
\begin{aligned}
    \nabla^2 H_{i,j} &= \frac{1}{6 (dx^2)} \big[ 4(H_{i+1,j}+H_{i-1,j}+H_{i,j+1}+H_{i,j-1})
    \\
    &\quad + H_{i+1,j+1}+H_{i-1,j+1}+H_{i+1,j-1}+H_{i-1,j-1}
    \\
    &\quad -20 H_{ij} + O(\nabla^4) \ .
\end{aligned}
\end{equation}
Similarly, following \cite{thampi-ronojoy}, the isotropic divergence of any vector ${\bf J}_{i,j}=(J^x_{i,j},J^y_{i,j})$ can be written as
\begin{equation}
\begin{aligned}
    \nabla \cdot {\bf J}_{i,j} &= \frac{3}{9 (dx)}\big[J^x_{i+1,j} - J^x_{i-1,j}+ J^y_{i,j+1} - J^y_{i,j-1}\big]
    \\
    &\quad + \frac{3}{36 (dx)} \big[ J^x_{i+1,j+1} + J^y_{i+1,j+1}+ J^x_{i+1,j-1}
    \\
    &\quad +J^y_{i-1,j+1} - J^x_{i-1,j+1} - J^x_{i-1,j-1} - J^y_{i-1,j-1}
    \\
    &\quad -J^y_{i+1,j-1} \big] + O(\nabla^3) \ .
\end{aligned}
\end{equation}
Specifically, such a choice corresponds to a finite difference scheme with a nine-point stencil that ensures both the isotropy condition and the proper relaxation to equilibrium (for the case $\omega=0$) on the lattice~\cite{LeVeque-book, thampi-ronojoy, Patra-Karttunen}.

For the instability analysis of the cycling state [Sec.~\ref{sec:floquet}], we integrate the dynamics of the homogeneous periodic phase $\Psi(t)$ [Eq.~\eqref{dynamical-psi}] using a Euler scheme with time step $\Delta t= 0.001$, and compute the corresponding monodromy matrices using the same temporal resolution, for a regular grid $q=(\Delta q, \ldots, 1)$ with $\Delta q=0.01$. 


\bibliography{references-dpm}

\end{document}